**Room Temperature Collective Blinking and Photon Bunching from $CsPbBr_3$ Quantum Dot Superlattice**

Qiwen Tan[1,5], Sudipta Seth[2], Boris Louis[2], Xiayan Wu[1,3], Nithin Pathoor[1], Toranosuke Takagi[1], Shun Omagari[1], Takumi Sannomiya[5], Johan Hofkens[2,4], Martin Vacha[1*]

[1]Department of Materials Science and Engineering, School of Materials and Chemical Technology, Institute of Science Tokyo, Ookayama 2-12-1, Meguro-ku, Tokyo 152-8552, Japan
[2]Laboratory for Photochemistry and Spectroscopy, Division for Molecular Imaging and Photonics, Department of Chemistry, Katholieke Universiteit Leuven, Leuven 3001, Belgium
[3]Department of Physical Science and Technology, Lingnan Normal University, Zhanjiang 524048, China
[4]Max Planck Institute for Polymer Research, Mainz 55128, Germany
[5]Department of Materials Science and Engineering, School of Materials and Chemical Technology, Institute of Science Tokyo, 4259 Nagatsuta-cho, Midori-ku, Yokohama, Kanagawa 226-8501, Japan

## Abstract

Development of quantum systems supporting collective many-body states is crucial for advancement of quantum technologies, and perovskite quantum dots (QDs) have emerged as promising quantum light sources. However, photon bunching, a key signature of collective states, has been observed in perovskites only at cryogenic temperatures. Here, we report collective blinking and photon bunching in perovskite QD superlattices at room temperature. Sub-wavelength-sized $CsPbBr_3$ superlattices exhibit distinct two-level blinking, and demonstrate photon bunching with a degree of up to 3.9. Time-resolved photoluminescence and super-resolution imaging reveal long lifetime components, and emission spatially confined to regions tens of nanometers in size, observations consistent with long-range exciton migration to a localized energy trap within the superlattice. Power-dependent degree of bunching and analysis of the bunching dynamics point to biexciton–exciton cascade emission as the origin of photon bunching. These findings establish perovskite QD superlattices as a promising platform for room-temperature collective optical phenomena.

## Introduction

Quantum technologies, such as computing, communications or sensing, have the potential to tackle some of the most challenging problems facing science and industry today, and pave the way for future innovations. To realize such technologies, there is a growing demand for developing quantum light sources and finding quantum systems capable of supporting collective many-body states. Such states, which are critical for understanding fundamental physics and for developing the new technologies, are typically realized at cryogenic temperatures. This restriction could be lifted by exploring new classes of materials, and quantum light sources have been demonstrated, e.g., using single organic molecules[1], single semiconductor quantum dots (QDs)[2] or single nitrogen-vacancy centers in diamonds[3]. Among the new materials, metal halide perovskite QDs are promising candidates due to their potential for non-classical light emission, such as single photons or entangled photon pairs. These non-classical states of light have been demonstrated in single perovskite QDs as photon antibunching[4-9], as well as photon bunching[10-13]. The latter phenomenon, which arises, among others, from cascade-like emission of biexcitons, further enhances the potential of perovskite QDs as quantum light sources, as the cascade emission is a key mechanism for generating entangled photon pairs[14-22]. Therefore, it is crucial to explore alternative systems that exhibit the correlated emission of entangled photons, as well as other collective phenomena at room temperature, with the vision to integrate perovskite QD systems into practical quantum technologies.

Recently, superlattices formed by cesium lead bromide perovskite ($CsPbBr_3$) QDs have emerged as a highly promising platform for investigating collective phenomena, arising from the interactions of individual quantum emitters. A typical $CsPbBr_3$ QD superlattice is fabricated via self-assembly of individual QDs into a periodic structure. The highly uniform stacking present in the superlattice creates an environment that enables coherent interactions between individual QDs. This gives rise to various types of collective behavior[23,24], such as the typically cryogenic phenomenon of superfluorescence[25]. Superfluorescence is preceded by the formation of a collective quantum state which results in shorter lifetime, red-shifted spectrum and photon bunching of the emission from the QD superlattice[25].

Cubic-shaped $CsPbBr_3$ QD superlattices can also act as optical cavities, with cooperative interactions between excitons and cavity modes, enabling lasing and cavity-enhanced superfluorescence[26-30]. In addition to collective optical behavior, collective charge dynamics have also been observed in $CsPbBr_3$ QD superlattices, leading to exciton delocalization and efficient exciton diffusion within the superlattice[29,31-33]. Beyond cubic shapes, superlattices with alternative shapes and building blocks have also been reported, and these exhibit unique characteristics as well. For example, didodecyldimethylammonium bromide (DDAB) treated $CsPbBr_3$ QDs can be used to form rhombic superlattices[34]. Low-dimensional lead halide perovskite superlattices[35] and anisotropic nanocrystal superlattices[36] exhibit good potential for light emitting diodes and show efficient carrier dynamics[37]. Superfluorescence has also been observed in some of these structures, including all-perovskite multicomponent superlattices[38], pyramid-shaped QD superlattices[39] and superlattices from perovskite nanocubes[40]. Although superfluorescence from perovskite QD superlattices is generally limited to cryogenic temperatures due to thermal decoherence[41], time-resolved spectroscopic features consistent with superfluorescence have been reported in perovskite thin films at higher[42] or even room temperature[43]. These results point to the great unexplored potential of perovskite hierarchical super-structures for collective optical phenomena at ambient temperatures.

Here, we investigate the photophysical properties of well-ordered QD superlattices at room temperature using single-particle. We focus on sub-wavelength size (100 ~ 500 nm) superlattices self-assembled from $CsPbBr_3$ QDs. We discovered collective blinking behavior and photon bunching at room temperature from individual superlattices, and propose exciton migration, biexciton formation and biexciton-exciton cascade emission as the origin of the observed phenomena.

## Results

### Fabrication and characterization of $CsPbBr_3$ quantum dot superlattices

$CsPbBr_3$ QDs with edge size of approximately 10 nm were synthesized using a modification of a previously reported room-temperature method[44]. TEM images of the synthesized QDs are shown in Supplementary Fig. 1. $CsPbBr_3$ QD superlattices were prepared on glass substrates by a self-assembly method, as shown in Fig. 1a. The QD toluene solution was spread onto the glass substrate and placed in a toluene-rich atmosphere, allowing the toluene to evaporate slowly and the QDs to gradually self-assemble into superlattices on the microscope cover glass. SEM images (Fig. 1b,c and Supplementary Fig. 2) reveal self-assembled cubic-shaped particles with edge lengths ranging from 100 nm to 500 nm. This size range is on the order of the wavelength of light and is significantly smaller than the several micrometer-sized perovskite superlattices reported previously[25-31]. About 50% of the samples exhibited irregular aggregates formed by stacking of the cubic-shaped particles. Some of these are also shown in the Supplementary Fig. 2. However, SEM alone cannot provide sufficient resolution to determine the precise internal structure of these cubic-shaped particles. This leaves two possibilities: these particles could be either superlattices assembled from individual QDs or, less likely, they might be big single crystals formed by the fusion of multiple QDs. To distinguish the two cases, we studied individual particles sufficiently dispersed on a substrate (as shown in Supplementary Fig. 3) using a fluorescence microscope. We note that the large majority of the particles show no signs of photobleaching for the period of the experiments (> 10 min.). We compared the PL spectra of an individual particle and of a random ensemble of isolated QDs. As shown in Fig. 1d, the spectrum of the QD ensemble exhibits a width (FWHM) of approximately 25 nm with a peak centered at 515 nm. In contrast, the superlattice particle shows a narrower FWHM of 15 nm, while the peak position remains at 513 nm. This reduction in FWHM and consistent peak position suggest that the particles are indeed superlattices in which individual QDs retain their quantum-confined character. A large fused single crystal would undergo significant spectral red-shift compared to the QDs because of the diminished quantum confinement effect[45-47], but unlike the red shifted spectrum associated with superfluorescence the bandwidth of the fused crystal would be broader. Since QDs with similar sizes are more likely to assemble into superlattices the spectrum is narrower due to the decreased size dispersion.

**Collective blinking and photon bunching at room temperature**

The PL characteristics of individual sub-wavelength size superlattices were measured at room temperature. Fig. 2a presents the time trace of PL intensity for a single superlattice and the corresponding PL intensity distribution. We examined such PL intensity traces of 202 superlattices and found that 193 of them (more than 95 %) exhibit unusual PL fluctuations, including two-level blinking behavior similar to single QDs, with the PL intensity switching between a strongly emitting ON-state and a grey state. To discriminate between the grey state and a possible OFF state we used also the PL lifetime and spectra, as elaborated later. Compared to the blinking of a single constituent $CsPbBr_3$ QD (Supplementary Fig. 4), the superlattices exhibit unique features in their blinking behavior. Specifically, the intensity of the ON-state is more than 20 times higher than that of the grey state, which itself is still weakly emissive. Also, under the same excitation conditions, the PL intensity of the ON-state of the superlattices is typically over 100 times larger than that of a single QD. These results suggest that the superlattice undergoes collective behavior, where the excitons within the superlattice transition synchronously between ON-states and grey states. For a reference, random QD ensembles of comparable sizes do not show any blinking due to averaging over many randomly blinking individual QDs[48] (Supplementary Fig. 5a).

We further carried out photon coincidence measurements using a 375 nm ps pulsed laser and the Hanbury-Brown Twiss detection scheme. As demonstrated in the second-order correlation function $g^{(2)}(t)$ (Fig. 2b) for the blinking superlattice, the photons of the ON-state are significantly bunched. The presence of bunching does not depend on the repetition rate within the range of 1 x $10^6$ ~ 1 x $10^7$ Hz. An example of photon bunching obtained with the repetition rate of 1 MHz is shown in Supplementary Fig. 6. For a statistical ensemble of 49 superlattices, the degree of bunching $g^{(2)}(0)/g^{(2)}(\tau)$ can reach up to 3.9 (Supplementary Fig. 8) and is anticorrelated with the superlattice PL intensity obtained under the same excitation intensity (Fig. 2c and Supplementary Fig. 8). Interestingly, while many superlattices display clear two-level blinking, some particles show blinking with more than two levels. For example, the superlattice illustrated in Supplementary Fig. 9a exhibits four-level blinking, characterized by three ON-states and one grey state. Photon coincidence measurements (Supplementary Fig. 9b) reveal that the total PL from such multi-level blinking particle is not bunched but the PL from the weakest ON-state remains bunched. This behavior suggests that such particles could be non-uniform aggregates (such as those in the bottom row of Supplementary Fig. 2) where each unit acts as an individual emitting center, and where only PL originating from specific uniform cubic region exhibits bunching. Another important observation is that the blinking and photon bunching are independent phenomena. High degree of photon bunching can be observed even for superlattices that do not show clear blinking, such as the example in Supplementary Fig. 10. On the other hand, no bunching is observed for random QD ensembles (Supplementary Fig. 5b), and single QDs show photon anti-bunching, as expected for a single photon emitter (Supplementary Fig. 11). We verified throughout the experiments that the phenomena we are observing are not artefacts originating from laser scattering or substrate emission. We note that the observations of anti-bunching from a single $CsPbBr_3$ QD and of bunching from a superlattice of such QDs are not necessarily contradictory because of the superlattice size and high local concentration of excitons. Such transition from antibunching to bunching were observed for other systems as well[49,50].

Fig. 2d shows the PL lifetimes of the ON-state (blue) and the grey state (black) for an individual superlattice. Both states display two-component PL lifetimes, with 17 ns and 45 ns components for the ON-state, and 4.6 ns and 22 ns components for the grey state. The longer components of the ON-state lifetimes measured on other individual superlattices are distributed between 28 and 69 ns, as shown in a histogram in Supplementary Fig. 12. The ON-state lifetime is significantly longer than that of the random ensemble of isolated single QDs (Supplementary Fig. 13) which have components of 12 and 2.4 ns, suggesting that the superlattices possess unique exciton dynamics. The grey state lifetime is comparable to that of the random QD ensemble, and this fact also helped with classifying the particular PL intensity as the grey state (as opposed to an OFF state). Lifetime analysis obtained with different thresholding of the ON- and grey states (Supplementary Fig. 14) confirms the existence of the grey state and clearly distinguishes it from a potential OFF state.

## PL blinking mechanism

The phenomenon of blinking is typically observed in individual nano-sized quantum emitters. Apart from that, collective blinking has also been reported in micro-sized objects and clusters of individual emitters. Such collective blinking behavior arises from various mechanisms. For instance, correlated blinking observed in bilayer 2D semiconductor heterostructures is associated with intermittent and random interlayer carrier transfer[51]. In perovskite microcrystals, spatially correlated blinking has been attributed to photon recycling caused by the reabsorption of emitted light and the waveguiding effect[52]. Various trap models have been proposed to explain collective blinking in semiconductor quantum wires[53] and perovskite sub-micrometer crystals[54,55]. Additional mechanisms include charge carrier[47,56,57] or exciton migration [58-62], which are linked to collective blinking observed in perovskite microcrystals and films, J-aggregates, multichromophoric dendrimers or CdSe semiconducting nanoplatelets.

To understand the mechanism of the observed collective PL blinking in the superlattices, we conducted super-resolution localization analysis. Fig. 3a presents the time trace of PL intensity for a single superlattice over a span of 600 seconds. In this specific case, the superlattice mainly remains in the grey state and switches to the ON state for short durations of time. Different PL intensity levels are represented by distinct colors, with the highest intensity marked in red and the lowest one in blue. Fig. 3b maps the localization positions of PL corresponding to these intensity levels. High PL intensity levels are confined within a region of approximately 30 nm in diameter. More examples for different types of blinking (shown in Supplementary Fig. 15) confirm the spatially confined nature of the emitting center within 20 – 30 nm. Such area is significantly smaller than the edge length of the superlattices (100 ~ 500 nm). We note that the super-resolution analysis provides only the size of the emitting area, not its position within the superlattice. The observation of such small emitting area suggests that even though absorption is supposedly occurring with the same probability at any location of the superlattice, the emission of the excitons is concentrated within a small localized region. Most excitons generated within the superlattice are funneled to an energy minimum (defect) located in this confined region, which acts as the primary luminescent center during the ON state.

More insights into the blinking dynamics were obtained through power spectral density (PSD) analysis of the PL time trace (Fig. 3c). More examples of blinking traces and PSDs are shown in Supplementary Fig. 16. The frequency $f$ dependent PSDs are fit[63] using a stretched Lorentzian function

$$\mathrm{PSD}(f) = \frac{A}{1+(f/f_0)^{\beta}} \qquad (1)$$

which consists of a constant $A$ in the low frequency range and describes a power law at high frequencies. $\tau$ is a characteristic time scale of PL fluctuations and is related to the characteristic frequency $f_0$ of the cross-over from saturation to power law as

$$2\pi f_0 = 1/\tau \qquad (2)$$

The majority of the blinking traces examined show saturation at the lower frequencies and the value of $\beta$ above 1.3. Such stretched Lorentzian behavior suggests that a well-defined type of phototactivable quencher controls the collective blinking behavior. For $\beta$ above 1.5 (such as the one in Fig. 3c) a single two-level photoactivated quencher with a specific switching rate is responsible for the blinking. Such quencher is associated with the spatially confined emission center where most excitons are funneled to and where they collectively transit between the emitting ON- and the grey states, leading to the observed collective blinking behavior. Deviation of $\beta$ values toward 1 suggests participation of additional fast two-level quenching processes. About 20% of the blinking traces showed such power-law type behavior without the low-frequency saturation, indicating that multiple two-level photoactivated quenchers with a distribution of switching rates contribute to the collective blinking.

Further, the fluorescence lifetime intensity distribution (FLID) of an individual superlattice was analyzed to understand the nature of the quencher (Fig. 3d). The distribution reveals distinct averaged lifetimes associated with the ON and grey states, suggesting a type-A blinking, in which the ON state corresponds to a neutral excitonic state and the grey state arises from charged excitons (positive or negative trions)[64,65]. Example of blinking traces obtained for a series of excitation powers from 0.05 W/cm$^2$ to 5.0 W/cm$^2$ (Supplementary Fig. 17) show a trend of increase of grey states fraction with increasing power, providing a support for the proposed mechanism.

## Model of the collective phenomena: Long-range exciton migration and biexciton-exciton cascade emission

We propose that the collective PL phenomena observed in the superlattices, including collective blinking and photon bunching, can be attributed to long-distance exciton migration, biexciton formation and biexciton-exciton cascade emission. When multiple excitons are generated within a perovskite QD, biexcitons can form depending on the QD size and halide composition[66]. The presence of biexcitons is typically confirmed by the saturation of emission intensity with increasing number of excitons[67], and by a spectrum which is red-shifted compared to single exciton emission due to the biexciton binding energy. The Coulomb binding energy can be affected by excitation energy[68], size of the QD[69,70] and degree of charge-carrier confinement[71], and in $CsPbBr_3$ QDs ranges from 5 to 100 meV[66-77]. Relaxation via biexciton-exciton cascade emission, where sequential one-exciton recombination events occur, results in photon bunching[78-82]. The cascade emission is a competing process with biexciton Auger recombination but it has been reported that the cascade emission in QDs has relatively high quantum efficiency[75]. The correlated photon pairs have been shown to exhibit quantum entanglement[14-22], making this process a promising candidate for quantum applications.

To confirm the presence of biexciton-exciton cascade emission in the superlattices, we investigated the excitation power dependence of the photon bunching degree and analyzed the bunching dynamics. We further supported the model by excitation power dependence of the PL intensity and by analysis of the PL spectra. Biexciton cascade emission is generally characterized by a distinct power dependence of the degree of photon bunching. Such dependence has been first observed on GaAs quantum dots[78], and later confirmed on InGaAs QDs[81], as well as perovskites[10,50]. The decrease of bunching degree with increasing excitation power is caused by the fact that at high excitation intensities (high number of biexcitons) the individual photon emission (from biexciton to exciton) surpasses the correlated photon pair emission from the biexciton-exciton cascade. The Fig. 4a shows the effect of excitation power on the degree of photon bunching for an example of two superlattices. With increasing excitation power, the degree of bunching decreases. A qualitative model proposed before[78,81] to explain the phenomenon is introduced in the Supplementary Information, together with the equation used to fit the data in Fig. 4a. Further, an example of the correlation function $g^{(2)}(t)$ taken at the lowest excitation intensities (Supplementary Fig. 18) confirms the bunching degree decrease with increasing power. We note that the trend of the decrease of the degree of bunching distinguishes biexciton-exciton cascade emission from other possible origins of photon bunching, such as correlated multiexciton emission,[83] superfluorescence[25] and super-radiance based lasing[84]. In these phenomena, an opposite trend of increase of photon bunching degree with increasing power has been observed. These findings support the interpretation that the observed photon bunching is a consequence of the biexciton-exciton cascade emission, and that potential contributions from higher multiexciton states can be neglected.

The biexciton formation is a result of accumulation of excitons in the confined location of the superlattice, as schematically summarized in Fig. 5. When a superlattice absorbs light, single excitons are generated at random positions throughout the structure. These excitons can either directly emit light through spontaneous emission, or migrate (funnel) to a spatially confined emitting site that works as an energy trap. Due to the trapping nature of this site the local exciton concentration increases, enhancing the likelihood of biexciton formation. It has been reported[75] that in perovskite QDs the exciton-exciton interaction is attractive, facilitating the formation and stabilization of biexcitons.

The excitation power dependence of PL intensity in Fig. 4b supports this hypothesis. The initial slope of the dependence is linear for the superlattice, the random QD ensemble, and for individual QDs, indicating the absorption due to single excitons in all cases. The single excitons created randomly in different parts of the superlattice then migrate and accumulate in the energy trap where they form the biexcitons. The trap is associated with the photoactive quencher that causes the collective blinking of excitons and biexcitons localized in the trap. We estimate (as detailed in the Supplementary Information) that at the excitation power of 1 W/cm$^2$ on average $N = 0.0075$ excitons are generated in a single QD. This estimate, together with the random nature of the absorption across the superlattice, supports the initial linearity of the power dependence of the superlattice and confirms that mutiexcitons are not directly generated in this intensity regime. We note that all the experiments (except the power dependences) are carried out with the 1 W/cm$^2$ excitation intensity. For higher excitation intensities the superlattice PL intensity saturates and this saturation is not observed in the random ensemble. Saturation of biexciton emission has been reported before[67]. However, in our case the saturation is a consequence of the energy funneling rather than a specific feature of the biexciton emission. In the Fig. 4b the single QDs show onset of PL saturation around the 100 W/cm$^2$ intensities, corresponding to $N = 0.75$. For a typical superlattice containing 1000 individual QDs, the total number $N$ under 1 W/cm$^2$ excitation is 7.5. Assuming a realistic value of PLQY of 80%, this number would drop to $N \sim 6$, as 20% of the excitons would decay non-radiatively before reaching the emission center. Further, assuming that the size of the emission center of 20 – 30 nm would correspond to ~ 10 QDs would result in $N = 0.6$ per QD, which is on the order of the onset of saturation. On the other hand, such funneling is not present in the random QD ensemble where individual QDs emit independently and the linearity is maintained.

The PL lifetime data are consistent with this model. Both the ON state and the grey state exhibit lifetime components comparable to the PL lifetime of random ensemble of QDs, corresponding to excitons outside the spatially confined emitting site. We note that the short ns component (the 2.4 ns component in the QD ensemble) is negligible in the ON-state lifetime, pointing to reduced non-radiative relaxation. For the ON-state, an additional longer lifetime component is present, which corresponds to the long-distance migration of the excitons. If the exciton, instead of emitting photon at the location of absorption transfers its energy (by Förster mechanism) to a neighboring QD and this step is repeated many times before reaching the emitting site, the lifetime of emission gets longer, resulting in the tens of ns lifetime component. On the other hand, biexcitons generally show very short PL lifetime (tens to hundreds of ps) which mostly reflects the Auger recombination process[67, 71, 77]. We have not resolved such short components in our experiments. However, in our photon correlation experiments we are concerned with the mechanism of cascade-like biexciton-exciton emission. Analysis of the photon bunching dynamics can, in principle, provide information on the cascade-like process[85]. Specifically, fitting of the central $g^{(2)}(\tau)$ peak and comparison with the fitting results of the side peaks enables extraction of a fast component which corresponds to the time delay between the emission of the first (biexciton) and second (exciton) photons in the biexciton-exciton cascade, and can be interpreted as the biexciton-exciton emission lifetime. We carried out such analysis on the $g^{(2)}(\tau)$ of several superlattices, and present an example in the Fig. 4c. The central peak is well fit with a two-exponential function with lifetimes of 1.3 ns and 27 ns, respectively. The side peaks show single exponential decay with a lifetime of 27 ns. We can therefore attribute the 1.3 ns component to the biexciton-exciton delay (or biexciton lifetime). We note that the 1.3 ns value is very likely resolution-limited, and that the actual delay is on sub-ns scales. This is consistent with the observation on $CsPbBr_3$ QDs that the radiative rate of the biexciton-exciton transition is more than two orders faster than that of the exciton transition[10].

Another experimental observation indirectly supporting this model is the dependence of the photon bunching degree on the PL intensity at the same excitation power (Fig. 2c) for different individual superlattices. The dependence shows an exponential-like decrease of the bunching degree with the PL intensity. Assuming that the PL intensity is directly related to the superlattice size, this result suggests that the larger the superlattice the smaller the degree of photon bunching. This might be a consequence of higher absolute number of excitons created by absorption in the larger superlattice which, when funneled from the whole superlattice to the low-energy trap site, leads to larger exciton density in the confined emitting space. Such situation is effectively identical to an increase of excitation power in a smaller superlattice, where we observed exponential decrease of the degree of bunching and explained that using the biexciton cascade-like emission model described by the Eq. (3). The same model can thus also explain the exponential-like decrease observed in Fig. 2c. A schematic figure comparing both phenomena is presented as Supplementary Fig. 19.

The Fig. 4d displays the PL spectrum of a single superlattice. The shape of the spectrum is identical for both the ON state and the grey state in the blinking traces (Supplementary Fig. 20). This fact also helped with classifying the particular PL intensity as the grey state (as opposed to an OFF state). Further, time evolution of the PL spectrum (Supplementary Fig. 21) does not show any signs of spectral changes (spectral diffusion), similar to other observations on $CsPbBr_3$ QDs[48]. The PL spectrum in Fig. 4d can be deconvolved into two Gaussian bands: One centered at 512 nm with an FWHM of 15 nm (purple), and another one at 517 nm with an FWHM of 29 nm (green). The red curve is the measured spectrum and the dashed black curve is the sum of the two Gaussian peaks, which matches the measured spectrum very well. The spectra of all superlattices are characterized by the presence of the low-energy Gaussian band. We note that neither the random QD ensemble (Fig. 1d) nor of the single QDs show the low-energy shoulder. Examples of the single QD spectra in the Supplementary Fig. 22 are characterized by mostly Lorentzian lineshapes. One plausible explanation of the unique superlattice spectral lineshape might be to attribute the red-shifted peak at 517 nm to biexciton-to-exciton transitions[71,76,77], and the peak at 512 nm to single exciton-to-ground state transitions. The red-shift would reflect the biexciton binding energy, of 23 meV in this case. The red shift observed for other superlattices ranges between 10 and 29 meV (Supplementary Fig. 23), which is well within the range of binding energies reported for the $CsPbBr_3$ QD materials[66-77]. On the other hand, we acknowledge that the contribution of the low-energy shoulder is relatively high, and that other explanations of the PL lineshape might be possible, such as exciton-phonon coupling specific to the superlattices that is not observed for the constituting QDs.

## Discussion

In summary, we successfully fabricated sub-wavelength size $CsPbBr_3$ QD superlattices with edge lengths ranging from 100 nm to 500 nm via a self-assembly process. The formation of these superlattices was confirmed by SEM imaging and spectral characterization. On single-particle level, the superlattices exhibit remarkable collective behavior, including synchronized blinking and photon bunching. Upon irradiation, single excitons are formed in individual QDs across the entire superlattice, and migrate efficiently to a confined low-energy region from which most of the PL takes place. The region has a size of 20 ~ 30 nm, which corresponds to approximately 10 (or less) QDs. This emission center can be associated with a photoactive quencher, in which case all PL originating from the center is temporarily quenched, leading to the appearance of the collective blinking. At the same time, the confined nature of the emission center leads to a significant increase in exciton density (up to $N \sim 6$ per center, or $N \sim 0.6$ per QD) which, together with the attractive nature of exciton-exciton interaction, leads to the formation of biexcitons, either intra- or inter-QD. Biexcitons in perovskite QDs have significant probability of cascade-like biexciton-exciton emission, competing with the Auger relaxation. The cascade-like emission results in the observation of photon bunching, with bunching degree of up to 3.9. Increased exciton density can potentially lead to the formation of higher multiexcitons that also show photon bunching. To distinguish the biexciton from multiexciton emission, as well as from other phenomena leading to photon bunching, we carried out excitation power dependence measurement of the bunching degree. The decrease of the bunching degree with increasing power is a signature of the biexciton cascade-like emission, and has qualitatively explained by simple theoretical models. Further support for the mechanism comes from dependence of the bunching degree on the PL intensity, and from the analysis of the bunching dynamics. Thus, both phenomena of collective blinking and photon bunching are intricately related, and are manifestations and consequences of the antenna effect of the whole superlattice and of funneling of the excitons to the confined emission center. These findings not only advance our understanding of exciton interactions within quantum dot superlattices but also provide valuable insights for the development of next-generation quantum and optoelectronic materials.

## Methods

*Synthesis of $CsPbBr_3$ quantum dots*

$CsPbBr_3$ quantum dots (QDs) are synthesized using a modified room-temperature synthesis procedure based on a previously reported method[40]. In a typical synthesis, 0.08 mmol CsBr and 0.08 mmol $PbBr_2$ were mixed with 2 mL of N,N-dimethylformamide (DMF). After stirring for 45 minutes, 200 μL of oleic acid (OA) and 100 μL of oleylamine (OAm) were added sequentially to stabilize the precursor solution. A total of 650 μL of the precursor solution was then added to 10 mL of toluene and stirred for 3 minutes. To purify the sample and minimize the size distribution of the QDs, the solution was centrifuged at 12,000 rpm for 10 minutes. After centrifugation, the supernatant was discarded, and the precipitate was re-dissolved in toluene for a second centrifugation at 12,000 rpm for 10 minutes. Finally, the precipitate was discarded, and the supernatant was collected for the fabrication of superlattices.

*Fabrication of the superlattices*

$CsPbBr_3$ quantum dot superlattices were prepared on 2.4 cm × 2.4 cm glass substrates by a self-assembly method. The glass substrates were first cleaned through a series of ultrasonic treatments: 5 minutes in acetone, followed by 5 minutes in ethanol, and finally 5 minutes in deionized water. After cleaning, a glass Petri dish was filled with toluene, and a glass stage approximately 3 cm × 3 cm × 2 cm in size was placed inside. The cleaned glass substrate was positioned on stage and 150 μL of the quantum dot dispersion in toluene at the concentration of $2.29 \times 10^{-7}$ mol/L was spread onto the glass substrate. The Petri dish was then covered with a glass lid to allow the toluene to evaporate slowly. During the evaporation of toluene, the quantum dots gradually self-assembled into superlattices on the glass substrate.

*Experimental setup*

The imaging, PL spectra and intensity fluctuation measurements were carried out using a wide-field inverted fluorescence microscope (IX 71, Olympus) with dry objective lens (UMPlanFl 100×, N.A. 0.95, Olympus). The samples were excited by a continuous-wave 360 nm laser (UV-FN-360-100 mW-5%-LED, CNI Laser) and the signal was detected with an electron-multiplying (EM) charge-coupled device (CCD) camera (iXon, Andor Technology). A 373 nm dichroic mirror and a 380 nm long pass filter were used to reject the excitation light. The excitation intensity was kept at 1 $W/cm^2$ for all experiments, except for the excitation intensity dependence measurements. For the spectral measurements, the PL signal was further dispersed by using an imaging spectrograph (CLP-50LD, Bunkou Keiki) placed before the EM-CCD camera.

The blinking trajectories were obtained with integration times of 30 ms. The trajectories were constructed by identifying a superlattice emission spot in the image, cropping a small area around the spot and integrating the PL intensity over the area. Area of the same size was then cropped and integrated near the superlattice spot where there was no emission visible, and this area was used as a background that was subtracted from the integrated superlattice signal. This process was repeated for all images in the series and the background-subtracted intensity was plotted as the blinking trajectory.

The time resolved PL measurements were carried out using the same microscope in a confocal mode, the Hanbury-Brown Twiss detection scheme and a time-correlated single photon counting (TCSPC) system (Pico Quant HydraHarp 400). The samples were excited with a picosecond 375 nm laser (LDH-D-C375, Pico Quant) focused to approximately 500 nm spot with the excitation intensity of 1 $W/cm^2$ and repetition rate within the range of 1 x $10^6$ ~ 1 x $10^7$ Hz. The PL signal from the samples was sent through a non-polarizing 50/50 beam splitter to two avalanche photodiodes (APD, Excelitas SPCM-AQRH-14). A 435 nm dichroic mirror and a 400 nm long pass filter were used to reject the excitation light. The arrival times of the photons relative to the laboratory time and the laser pulse time were recorded, and the PL time trajectory, the PL decay curve and the second-order correlation $g^{(2)}(t)$ function were constructed. The resolution limiting factor in this setup is the APD jitter (uncertainty in the photon-to-electron conversion). Such timing resolution for the SPCM-AQRH-14 is product-specified as 225 ps. However, the real resolution is longer, on the sub-ns order. The degree of bunching was calculated based on the ratio of peak intensities after background subtraction. Representative data were also analyzed by integrating the peak areas, and presented in the Supplementary Fig. 8.

The blinking trajectories were acquired using the TTTR acquisition mode, and the HydraHarp software used for the analysis enables integration times (binning) as short as 0.2 ms. To determine the appropriate binning time, we analyzed the blinking traces using a series of binning times between 0.2 ms and 10 ms. Two such examples are shown in the Supplementary Fig. 24 and Supplementary Fig. 25. With the shorter times (< 2 ms), the individual intensity levels in the histogram get smeared out due to the increased noise. At the same time, shortening of the binning time does reveal any additional features such as unresolved OFF or grey states. We therefore used the 10 ms bin as the optimum compromise between the S/N ratio and the resolution. To acquire the background, identical measurements were carried out from nearby locations without detectable emission. All blinking trajectories presented are background-subtracted in this way.

*Super-resolution localization analysis*

Super-resolution localization methods typically rely on the stochastic switching of individual emitters and determining their centroid positions in the photoluminescence (PL) image. A diffraction-limited PL spot, whose size is similar to or slightly larger than the point spread function (PSF) of the microscope, may contain multiple nanoscale emitters. To determine the shape and central location of the emission profile, a rotational 2D Gaussian function is commonly used for fitting:

$$G(x, y, \theta) = A\ exp\ (-\ (a(\mu_x - x_0)^2 + 2b(\mu_x - x_0)(\mu_y - y_0) + c(\mu_y - y_0)^2))$$

where

$a = (cos^2\theta)\ /\ (2\sigma_x^2) + (sin^2\theta)\ /\ (2\sigma_y^2)$, $b = (sin2\theta)\ /\ (4\sigma_x^2) + (sin2\theta)\ /\ (4\sigma_y^2)$,
$c = (sin^2\theta)\ /\ (2\sigma_x^2) + (cos^2\theta)\ /\ (2\sigma_y^2)$.

Here, $A$ represents the amplitude, ($\mu_x$, $\mu_y$) denote the mean position in the $x$-$y$ plane, ($\sigma_x$, $\sigma_y$) correspond to the widths along the major and minor axes, and $\theta$ indicates the angle of rotation relative to a reference point in the $x$-$y$ plane. If multiple emitting objects exist within the diffraction-limited region and operate independently, the centroid position fluctuates over time depending on the PL-active and inactive states of the objects.

In the $CsPbBr_3$ system studied here, individual superlattice, aggregates or isolated nanocrystals may emit separately, especially when charge or energy transfer among them is minimal. By tracking the Gaussian centroid position across consecutive frames of a recorded sequence, one can determine the spatial emission coordinates, referred to as the localization position. The spatial distribution of these positions, along with intensity correlations, enables the identification of the number of emitting objects and their relative alignment. If a single localization position is detected within a spot, it suggests that only one emitter is present, within which excitons diffuse freely. Conversely, the presence of multiple localization positions implies multiple emission centers likely originating from more than one superlattices or aggregates to the overall PL signal.

*Power spectral density estimation*

The power spectral density (PSD) of blinking traces is estimated by applying the Fourier transform to the intensity autocorrelation function, following the Weiner-Khinchin theorem. Here, PSD models intensity fluctuations involving a two-level system within the framework of the multiple recombination center model[86]. Details of the method can be found elsewhere[63].

**DATA AVAILABILITY**

The data that support the findings of this study are available from the corresponding author upon request.

## ACKNOWLEDGMENTS

M.V. acknowledges the support by the JSPS KAKENHI grant number 24K01449 and by the JSPS KAKENHI grant number 23H04875 in Grant-in-Aid for Transformative Research Areas 'Materials Science of Meso-Hierarchy'. S.S. acknowledges the support of Marie Skłodowska-Curie postdoctoral fellowship (No. 101151427, SPS_Nano) from the European Union's Horizon Europe program, short stay abroad grant (K257023N) and travel grant (K147824N) from Research Foundation-Flanders (FWO). B.L. thanks FWO for his Junior Postdoctoral fellowship (12AGZ24N). J.H. acknowledges financial support from the Research Foundation-Flanders (FWO, Grant Nos. G098319N, G0F2322N, S002019N, VS06523N, G0AHQ25N) from the Flemish government through long-term structural funding Methusalem (CASAS2, Meth/15/04), and the MPIP as MPI fellow. Q.T. thanks W. He, X. Cao and Y. Wang for technical assistance. Q.T. is also grateful to X. Wang, F. Hu and Y. Tang for their fruitful discussions and suggestions.

## AUTHOR CONTRIBUTIONS

Q.T. synthesized the superlattices; Q.T., S.S, B.L., X.W., N.P., T.T., S.O. performed the experiments and analyzed the results; T.S. provided theoretical analysis; J.H. and M.V. supervised the research; Q.T. and M.V. wrote the paper.

## COMPETING INTERESTS

The authors declare no competing interests.

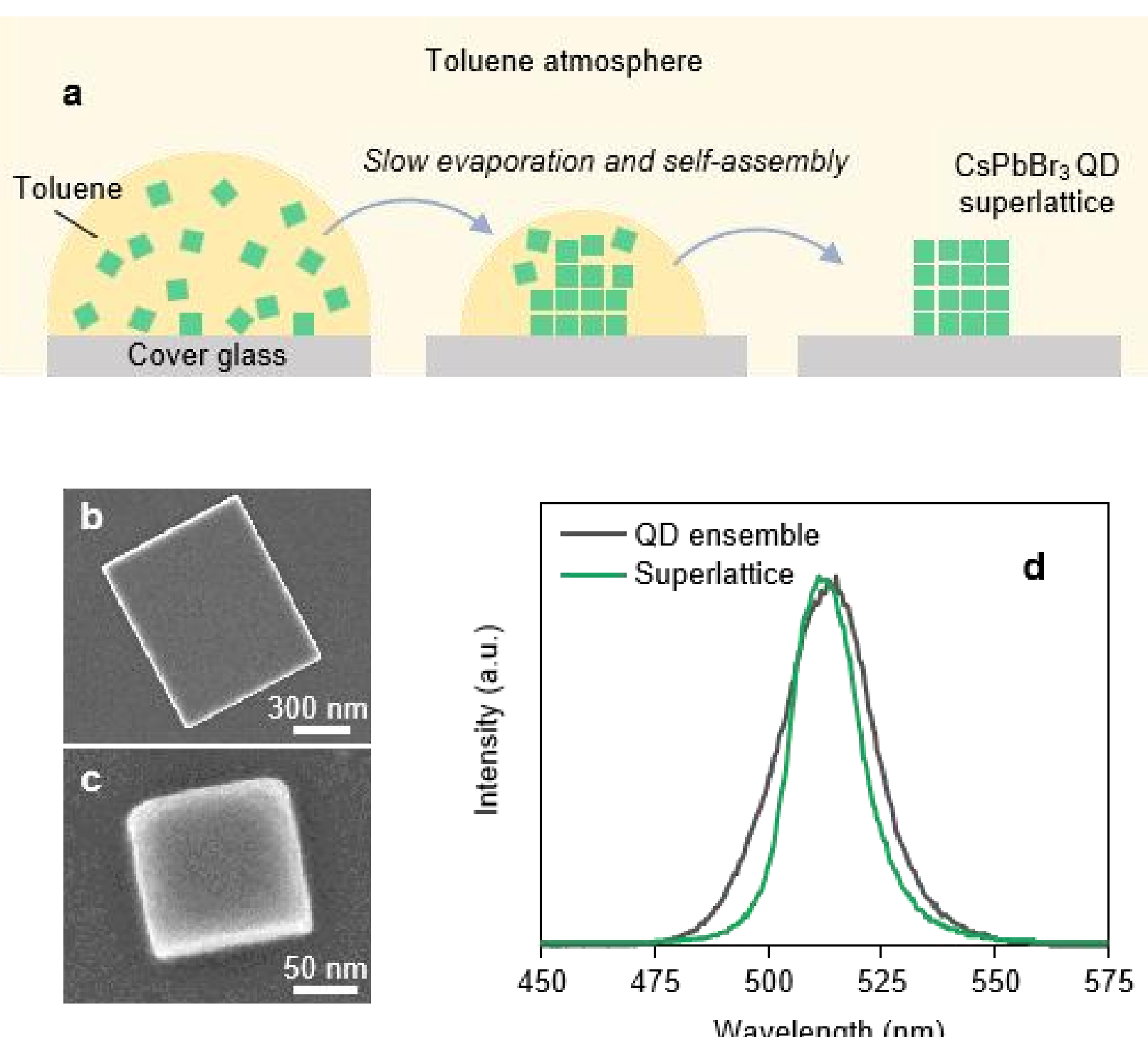

**Fig. 1** Preparation and characterization of the superlattices. (a) Scheme of the self-assembly of a $CsPbBr_3$ superlattice from individual $CsPbBr_3$ QDs in toluene; (b), (c) SEM images of typical cubic shaped $CsPbBr_3$ QD superlattices; (d) PL spectrum of an ensemble of $CsPbBr_3$ QDs (black) and of a single superlattice (green)

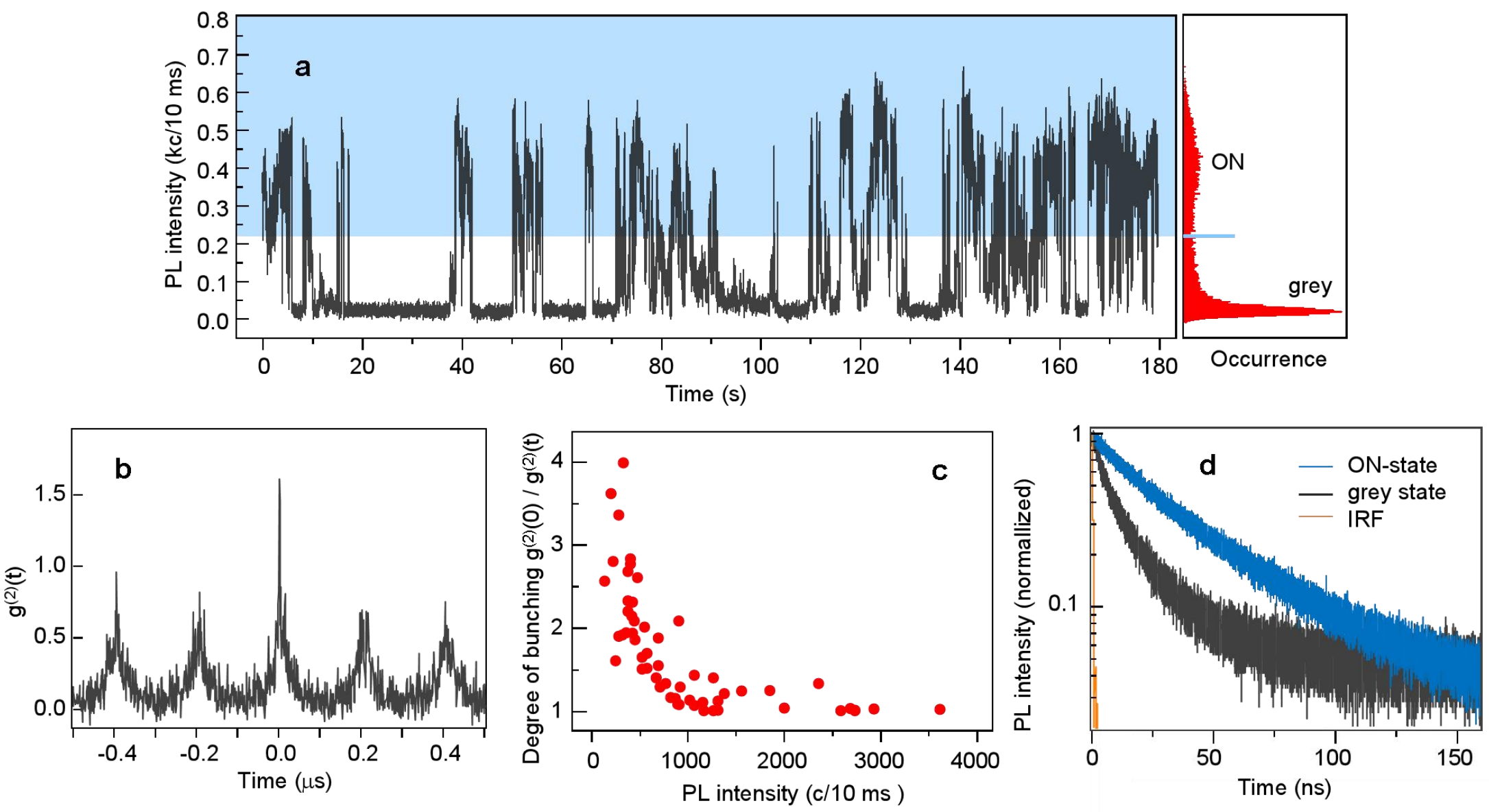

**Fig. 2** Collective phenomena in a single superlattice. (a) Left: PL intensity fluctuation of a $CsPbBr_3$ superlattice; the blue shaded area indicates the PL intensity range of the ON-state; Right: Histogram of PL intensity levels corresponding to the time trace; (b) Second-order correlation function measured on a single superlattice indicating high photon coincidence; (c) PL intensity dependent degree of photon bunching obtained with the same excitation power on 49 single superlattices; (d) PL lifetime of the ON-state (blue) and the grey-state (black) of the single superlattice in (a); the orange line represents the instrument response function. All data were acquired using the APD.

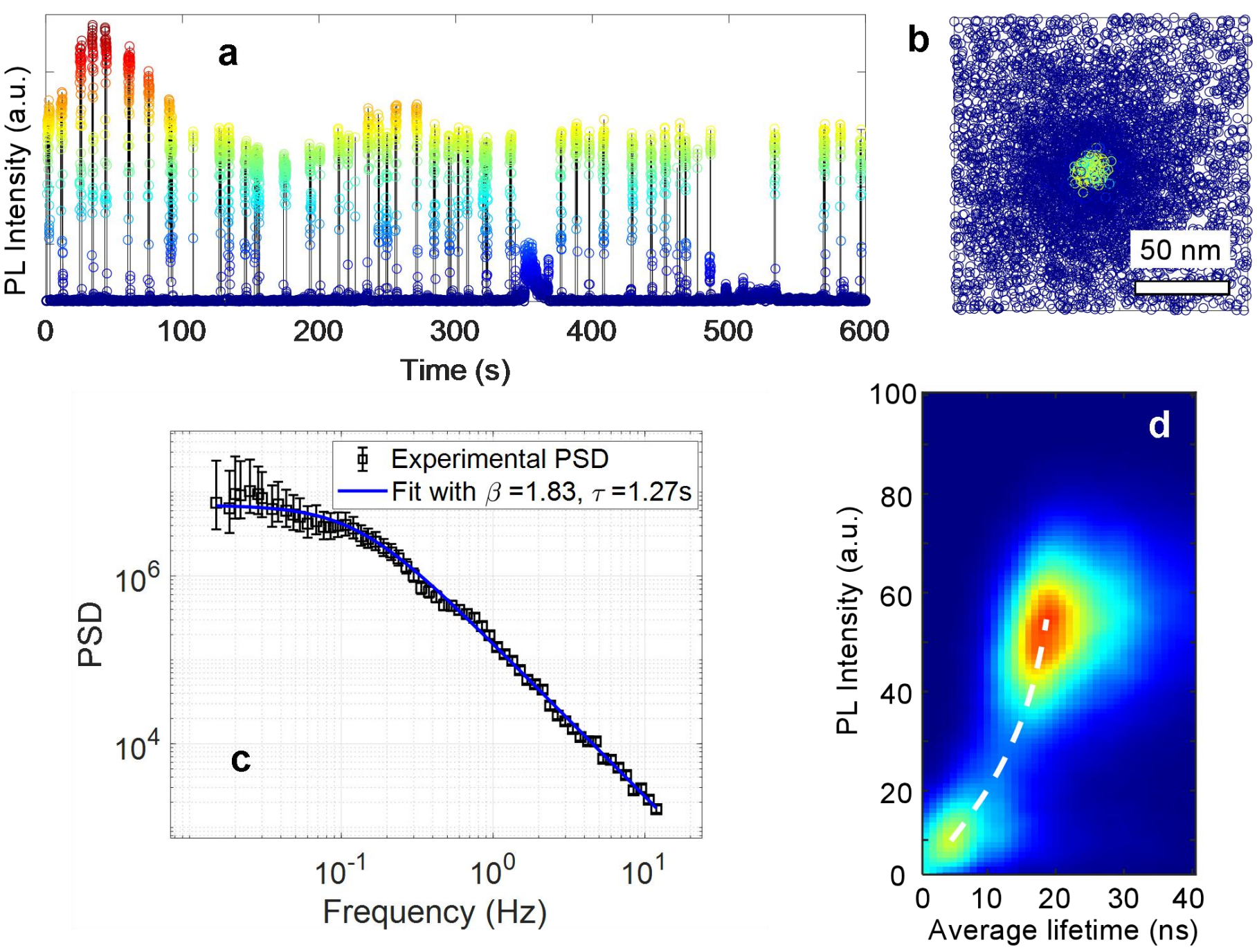

**Fig. 3** Collective blinking characteristics. (a) PL intensity fluctuation of a $CsPbBr_3$ superlattice; different PL intensity levels are represented by distinct colors, with the highest intensity marked in red and the lowest in blue; the data were acquired using EM-CCD; (b) Localized positions of PL corresponding to different intensity levels; (c) Power spectral density (PSD) of PL intensity fluctuation plotted in (a); the inset shows parameters obtained by fitting the PSD using the Eq. (1); (d) Fluorescence lifetime intensity distribution (FLID) of a single superlattice.

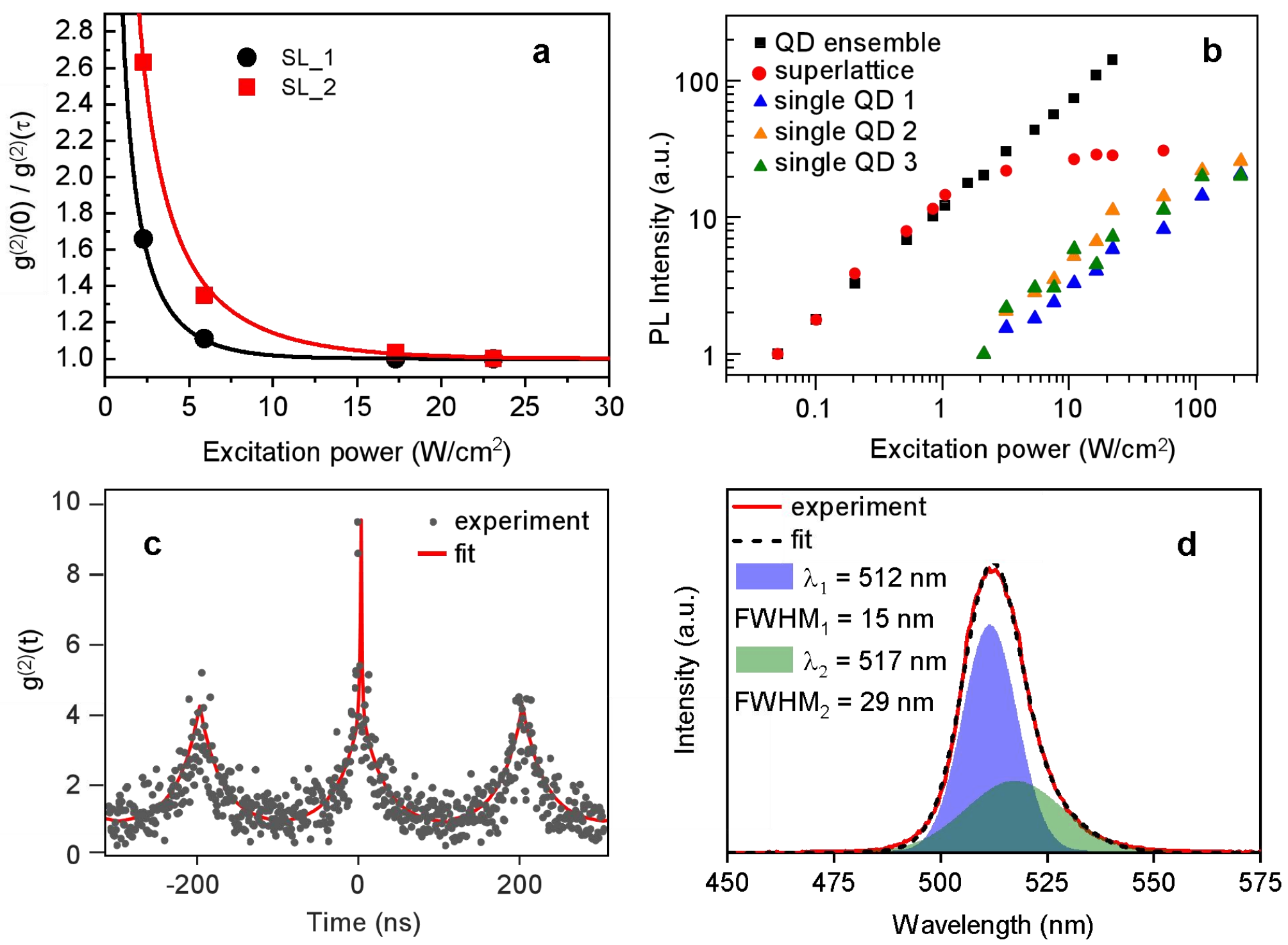

**Fig. 4** Analysis and characterization of photon bunching and bi-exciton emission. (a) Excitation power dependent degree of photon bunching of two different superlattices. The solid lines represent fits to the data as described in the Supplementary Information; (b) Excitation power dependent PL intensity of a random QD ensemble (black), a single superlattice (red) and single QDs (blue, orange and green); (c) Second-order correlation function measured on a single superlattice (symbols) and its fit with two-exponential function (solid line); (d) PL spectrum of a single superlattice (solid red line) fitted by two Gaussian peaks (dashed black line); the two components are shaded in purple and green.

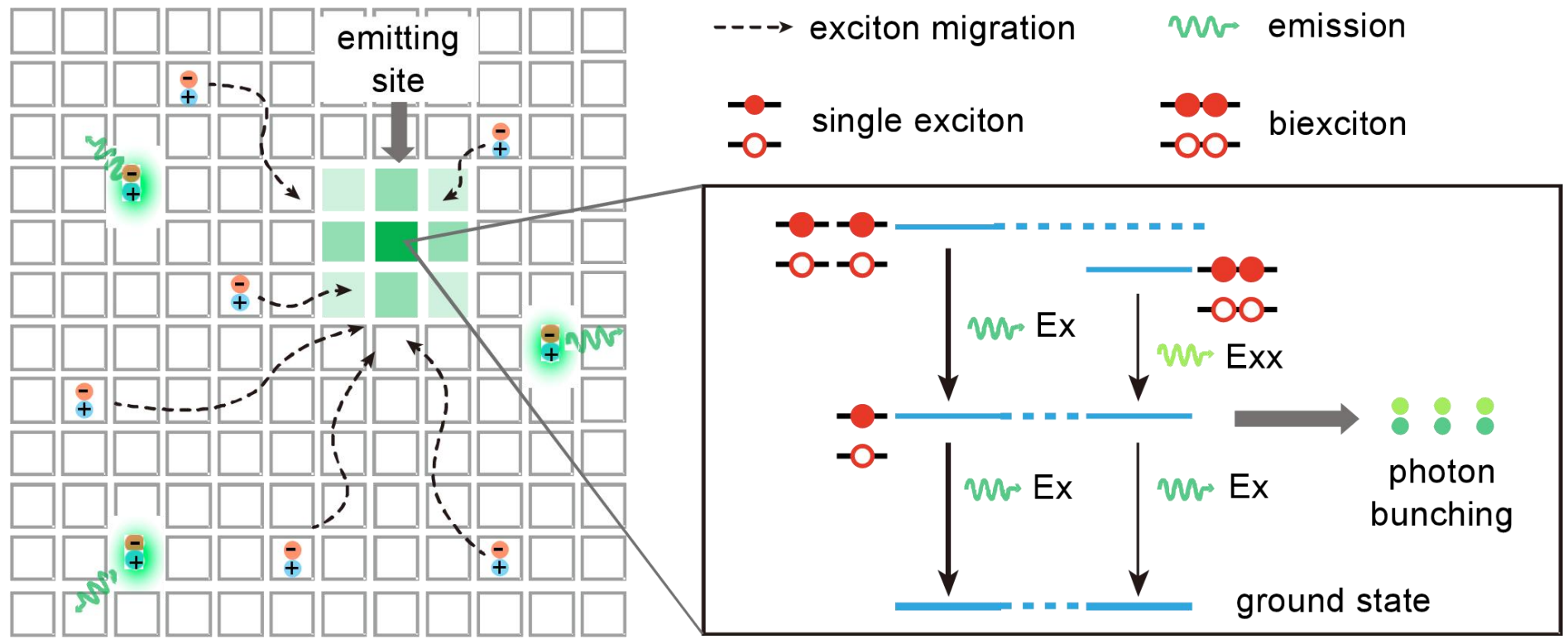

**Fig. 5** A schematic model of exciton migration, biexciton formation and biexciton-exciton cascade emission in a superlattice

**Supplementary Information**

Room Temperature Collective Blinking and Photon Bunching from $CsPbBr_3$ Quantum Dot Superlattice

Qiwen Tan[1,5], Sudipta Seth[2], Boris Louis[2], Xiayan Wu[1,3], Nithin Pathoor[1], Toranosuke Takagi[1], Shun Omagari[1], Takumi Sannomiya[5], Johan Hofkens[2,4], Martin Vacha[1*]

[1]Department of Materials Science and Engineering, School of Materials and Chemical Technology, Institute of Science Tokyo, Ookayama 2-12-1, Meguro-ku, Tokyo 152-8552, Japan
[2]Laboratory for Photochemistry and Spectroscopy, Division for Molecular Imaging and Photonics, Department of Chemistry, Katholieke Universiteit Leuven, Leuven 3001, Belgium
[3]Department of Physical Science and Technology, Lingnan Normal University, Zhanjiang 524048, China
[4]Max Planck Institute for Polymer Research, Mainz 55128, Germany
[5]Department of Materials Science and Engineering, School of Materials and Chemical Technology, Institute of Science Tokyo, 4259 Nagatsuta-cho, Midori-ku, Yokohama, Kanagawa 226-8501, Japan

**Fig. 18** Photon coincidence measurement of emission from the same superlattice at different excitation powers
**Fig. 19** Schematic explanation of the decrease of the degree of bunching
**Fig. 20** PL blinking trace and PL spectra of ON-states and grey-states of a single superlattice
**Fig. 21** Time evolution of a PL spectrum of a single superlattice
**Fig. 22** Examples of PL spectra of single $CsPbBr_3$ QDs
**Fig. 23** Histogram of the energy difference between the two components of deconvolved PL spectra
**Fig. 24** PL intensity fluctuation of a single blinking superlattice
**Fig. 25** PL intensity fluctuation of a single non-blinking superlattice

**Relationship between the number of excitons and the degree of photon bunching**

To qualitatively explain the experimental observation that the degree of bunching decreases with increasing number of excitons (excitation power), as shown in Fig. 4a, we introduce below a simple model presented in ref. [78] and [82]. We acknowledge that the model is based on many oversimplifying assumptions, such as that the presence of two excitons in the system at $t = 0$ results in simultaneous emission of both excitons. Still, the model it is capable of reproducing qualitatively the overall trend observed in the experiment.

If we assume that the number of excitons $n$ follows a Poisson distribution, then the probability that $n$ excitons exist is expressed as:

$$P_N^n = \frac{1}{e^N}\frac{N^n}{n!}$$

where $N$ is the mean number of excitons which is proportional to the power of the pulsed excitation laser. Suppose $P_X$, $P_{XX}$ are the probabilities that at least one exciton and that two excitons exist, respectively. Then, $P_X$ and $P_{XX}$ can be written as:

$$P_X = 1 - P_N^0$$
$$P_{XX} = 1 - P_N^0 - P_N^1$$

where

$$P_N^0 = \frac{1}{e^N}\frac{N^0}{0!} = \frac{1}{e^N}$$
$$P_N^1 = \frac{1}{e^N}\frac{N^1}{1!} = \frac{N}{e^N}$$

are the probabilities that 0 exciton and 1 exciton exist, respectively. If $\eta$ is the detection efficiency of the two single photon detectors of the Hanbury Brown Twiss setup. Then, $g^{(2)}(0)$, which is the photon coincidence at $t = 0$, is given by at least two excitons existing in the system and both photons emitted getting detected by the detectors:

$$g^{(2)}(0) = P_{XX} \cdot \eta \cdot \eta$$

The side peaks, which are the coincidences at $t = \pm\tau$, are given by two cases: 1) only one exciton is excited by the first laser pulse, and the emitted photon is detected by one of the detectors; and more than one exciton is excited by the second laser pulse and the photons emitted are detected by another detector; 2) at least two excitons are excited by the first laser pulse, but only one of the detector detects the photons emitted; then at least one excitons is excited by the next laser pulse and the other detector detects the photons emitted. Then, the intensity of side peaks $g^{(2)}(\tau)$ is given by:

$$g^{(2)}(\tau) = P_N^1 \cdot \eta \cdot P_X \cdot \eta + P_{XX} \cdot \eta \cdot (1-\eta) \cdot P_X \cdot \eta$$

Since the possibility to find only one exciton is relatively small, and the detection efficiency is very low, the degree of bunching is given by:

$$\frac{g^{(2)}(0)}{g^{(2)}(\tau)} = \frac{1}{P_X} = \frac{1}{1-P_N^0} = \frac{1}{1-e^{-N}}$$

where $g^{(2)}(0)/g^{(2)}(\tau)$ is the degree of bunching. This equation qualitatively describes the decrease of the bunching degree with increasing number of excitons (excitation power), and was used to fit the data in Fig. 4a.

Alternatively, to qualitatively describe the degrease of bunching with increasing excitation power, we could model the biexciton as a two-photon state, which can be approximated as

$$\rho = p_0|0\rangle\langle 0| + p_2|2\rangle\langle 2|$$

where always two photons exist when emitted. Since the denominator and nominator of

$$g^{(2)}(0) = \frac{\langle a^\dagger a^\dagger a a\rangle}{\langle a^\dagger a\rangle^2} = \frac{\langle n(n-1)\rangle}{\langle n\rangle^2}$$

can be expressed using

$$\langle n\rangle = \sum n p_n = 2p_2 \text{ and } \langle n(n-1)\rangle = \sum n(n-1)p_n = 2p_2$$

respectively, we obtain

$$g^{(2)}(0) = \frac{1}{2p_2}$$

The excitation power increase corresponds to the increase of $p_2$, thus reduces the bunching. We further assume a Poissonian distribution of excitons that all end up in two photons as biexcitons by reducing the number by a factor of 2,

$$p_2 = \frac{1}{2}\sum_{n=1}^{\infty} P_N^n = \frac{1}{2}(1-e^{-N}) \text{ with } p_0 \gg p_2 \text{ for the simplicity of normalization.}$$

This results in

$$g^{(2)}(0) \sim \frac{1}{1-e^{-N}}$$

which qualitatively describes the phenomenon in the same way as the above model.

**Estimation of number of excitons *N* generated by a laser pulse in a single quantum dot**

The number of excitons *N* generated in a single QD by a laser pulse is expressed as:

$$N = N_p . \sigma_{QD} \ ,$$

where *Np* is density of photons arriving at the QD and $\sigma_{QD}$ is the absorption cross section of the QD, which is estimated as $2x10^{-14}$ $cm^2$.
Density of photons *Np* is given by:

$$N_p = \frac{W_{laser}}{f_{rep} E_p}$$

where $W_{laser} = 100 \ \mathrm{W/cm^2}$ is the power of the laser, $f_{rep} = 5 \times 10^6$ Hz is the laser repetition rate and $E_p = 3.31$ eV is the energy of photon.
Therefore, the number of excitons generate per pulse by a laser with 100 $W/cm^2$ power and $5x10^6$ repetition rate is calculated as 0.75.

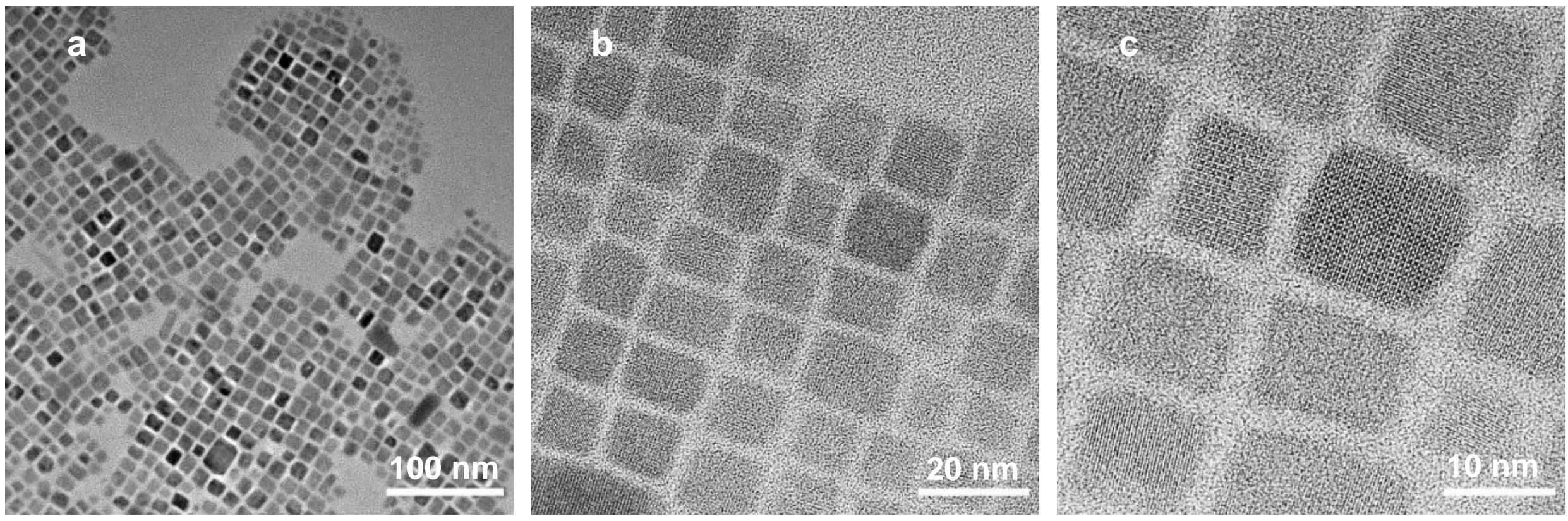

**Supplementary Fig. 1** TEM images of $CsPbBr_3$ quantum dots used for the superlattice synthesis. The QDs were synthesized using a room temperature method, with the resulting edge size of approximately 10 nm.

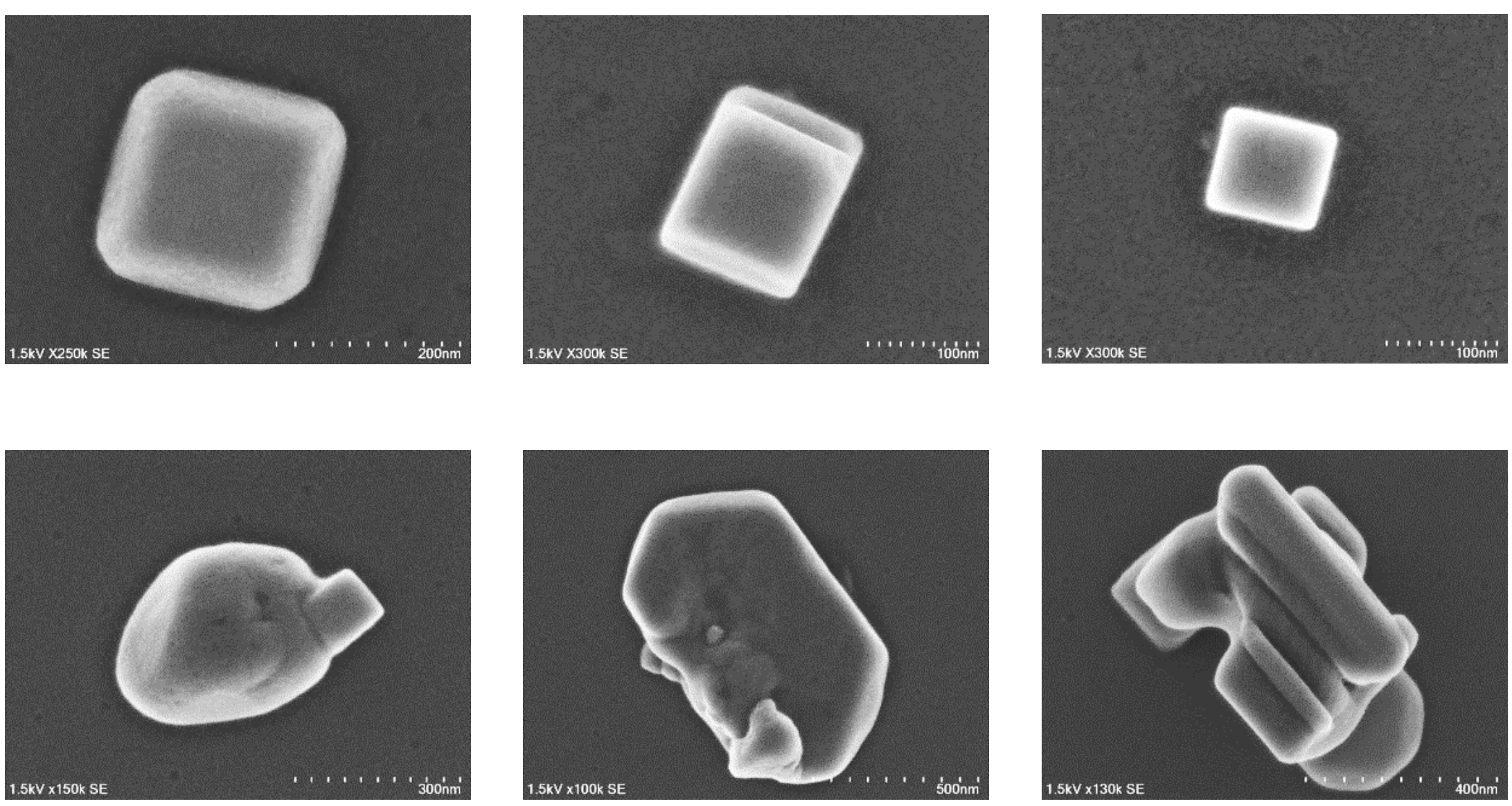

**Supplementary Fig. 2** Examples of SEM images of regular cubic superlattices (top row) and of superlattices with non-uniform shape (bottom row). The irregular shapes represent about 50% of all superlattices measured by the SEM.

The irregular-shaped superlattices are significantly larger that the regular ones, and as such emit with higher intensities. We assume that the complex shape also leads to multi-level blinking, and we used this criterion to include only the two-level blinking particles in the analysis. The maximum emission intensity of the multi-level blinking particles is about twice or more than that of the two-level blinking particles.

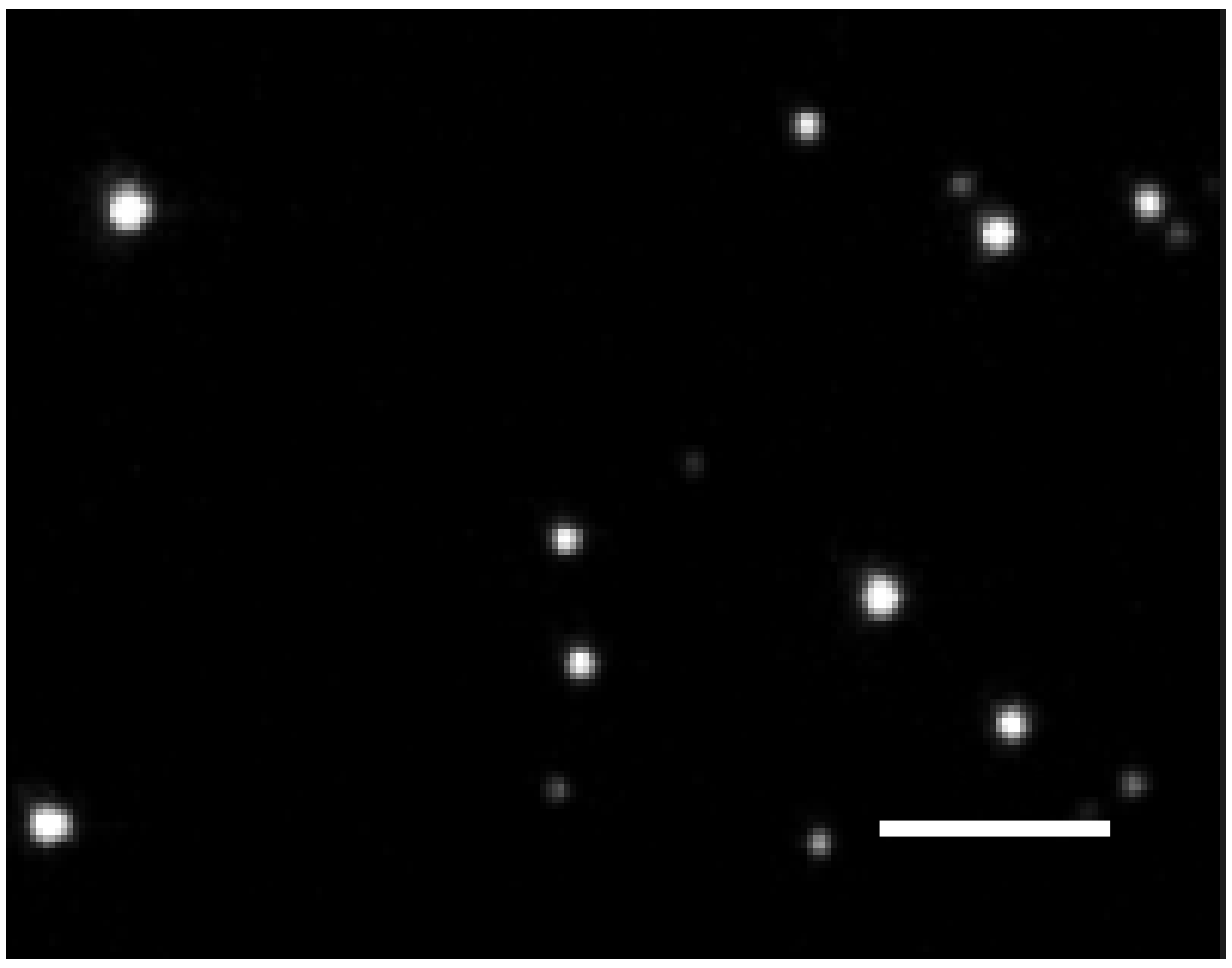

**Supplementary Fig. 3** Microscopic fluorescence image of individual QD superlattices dispersed on a glass substrate excited with a 360 nm laser. The image confirms sufficient spatial separation between the superlattices and shows distribution of the PL intensities from the individual emitting spots.

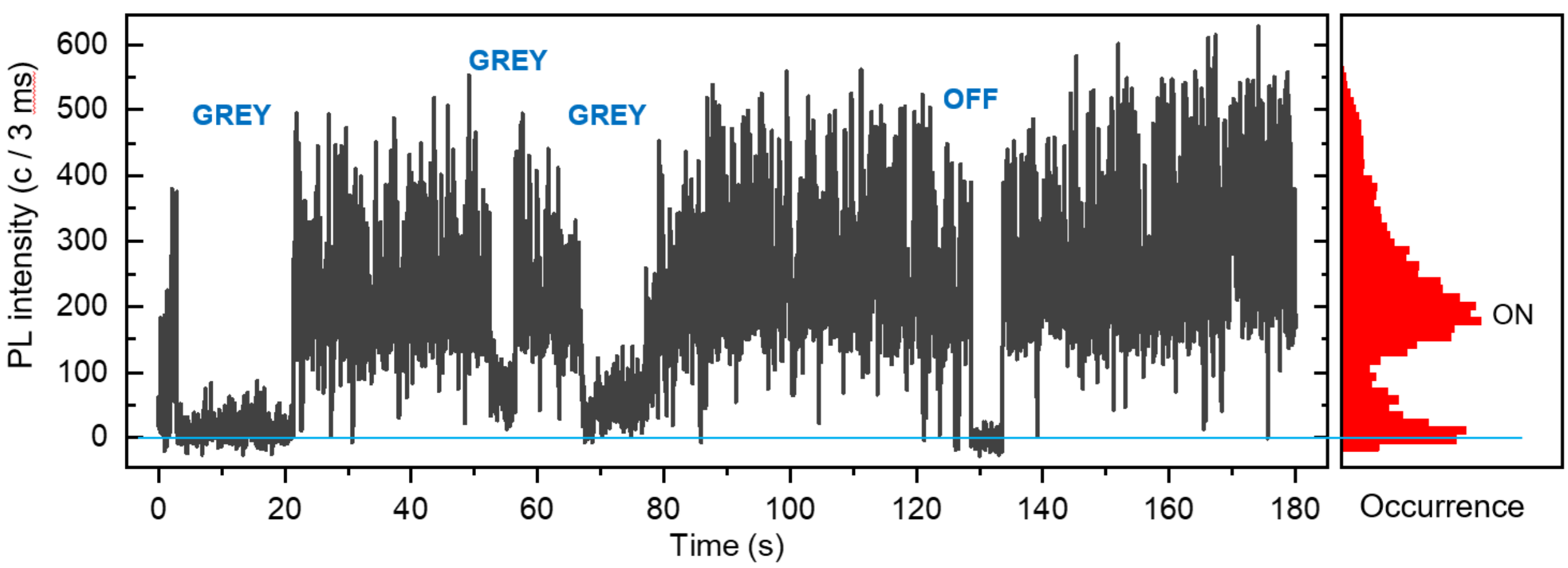

**Supplementary Fig. 4** PL Blinking trace of a single $CsPbBr_3$ quantum dot (left) and PL intensity histogram (right). This particular QD shows an ON-state intensity, several periods of grey states, and an OFF-state around 130 s. The states are indicated in the figure. The data were acquired using EM-CCD.

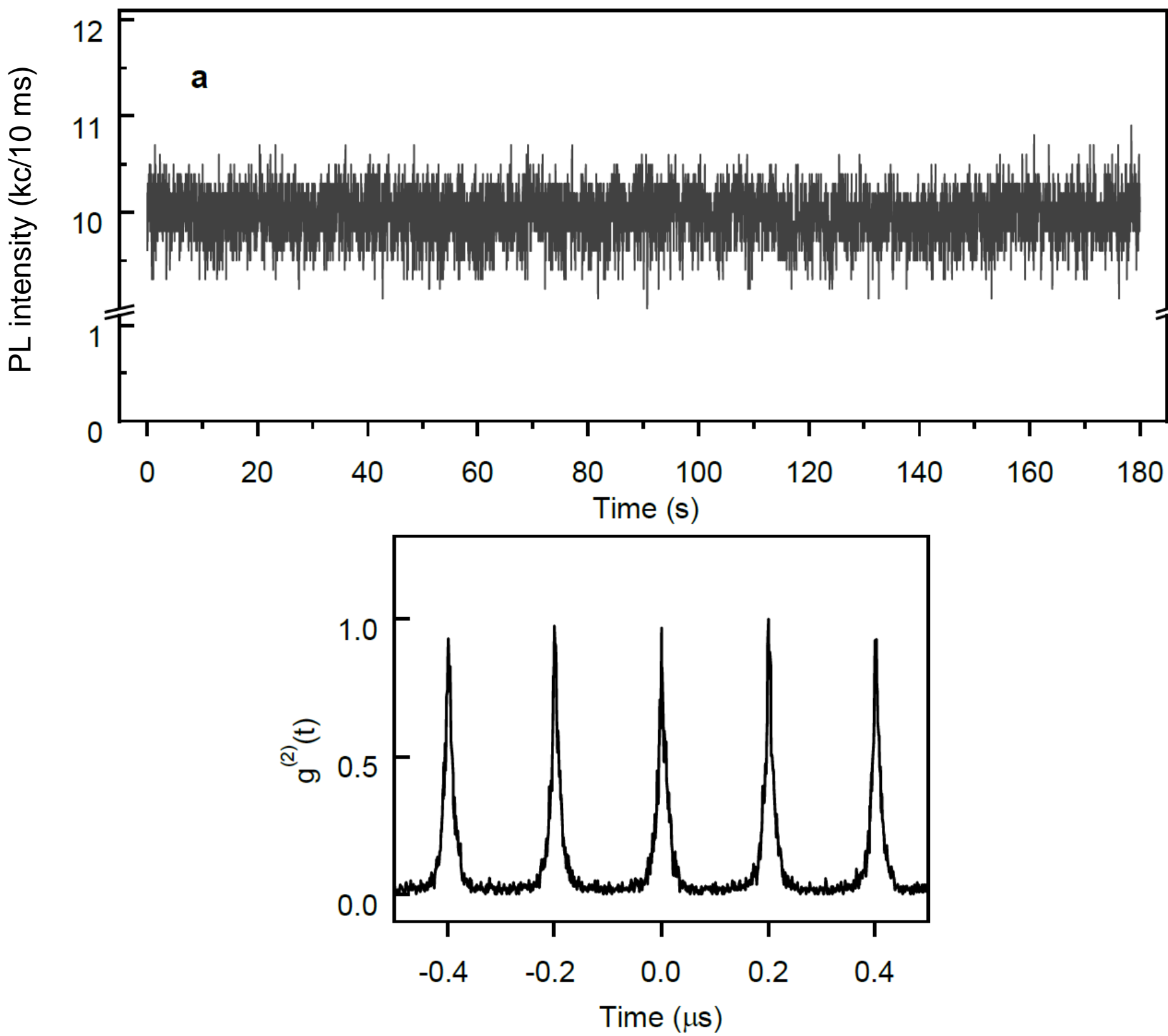

**Supplementary Fig. 5** (a) PL intensity fluctuation of a random ensemble of $CsPbBr_3$ QDs; (b) Photon coincidence measurement of emission from a random ensemble of $CsPbBr_3$ QDs. No photon bunching is evident from correlation function plot.

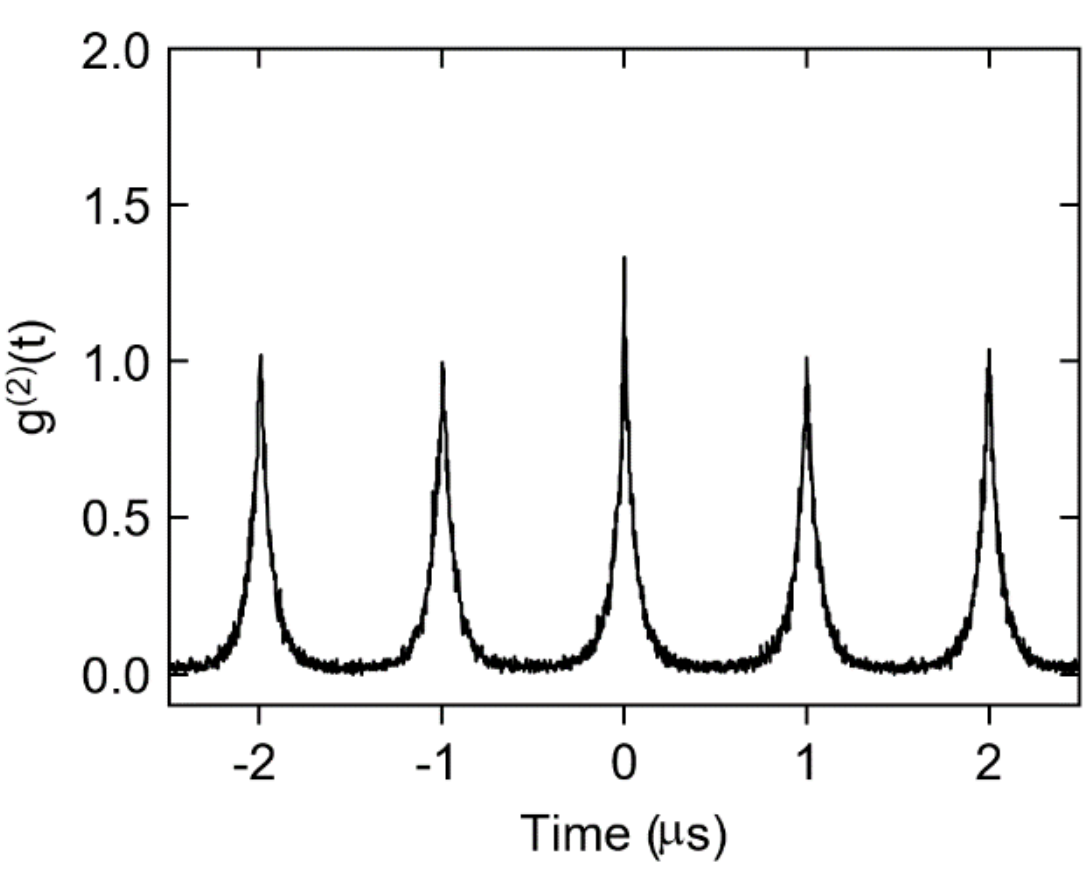

**Supplementary Fig. 6** Photon coincidence measurement of emission from a single $CsPbBr_3$ QD superlattice using the repetition rate of 1 MHz. This particular superlattice showed photon bunching with a degree of 1.3.

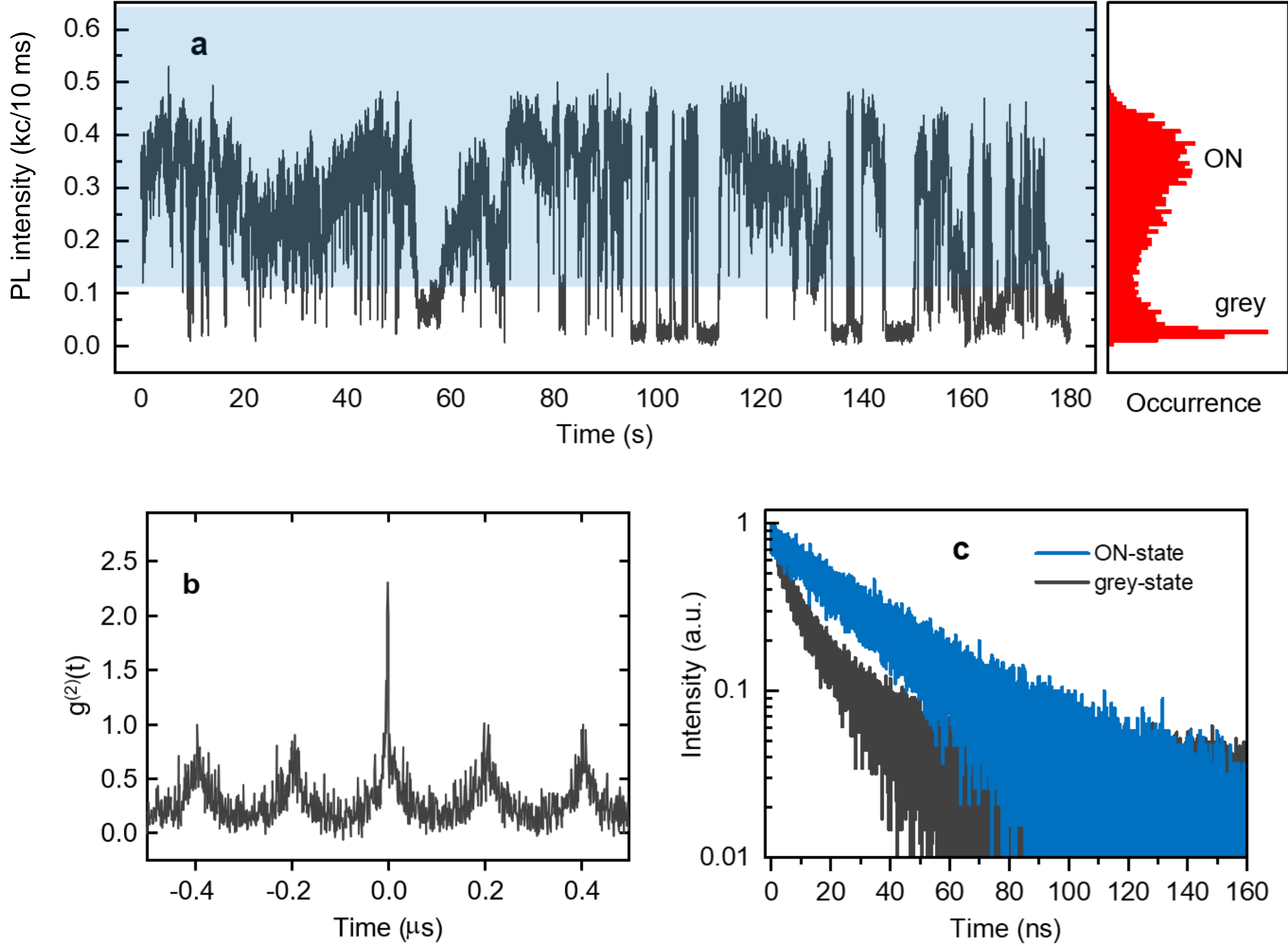

**Supplementary Fig. 7** (a) Left: PL intensity fluctuation of a single superlattice; the blue shaded area indicates PL intensity classified as the ON-state; Right: Histogram of PL intensity levels during the blinking time interval; the ON-state and grey-state levels are indicated in the figure; (b) Photon coincidence measurement of emission from the same superlattice, showing photon bunching with a degree of 2.3; (c) PL lifetime of ON-state (blue) and grey-state (black) of the superlattice. All data were acquired using the APD.

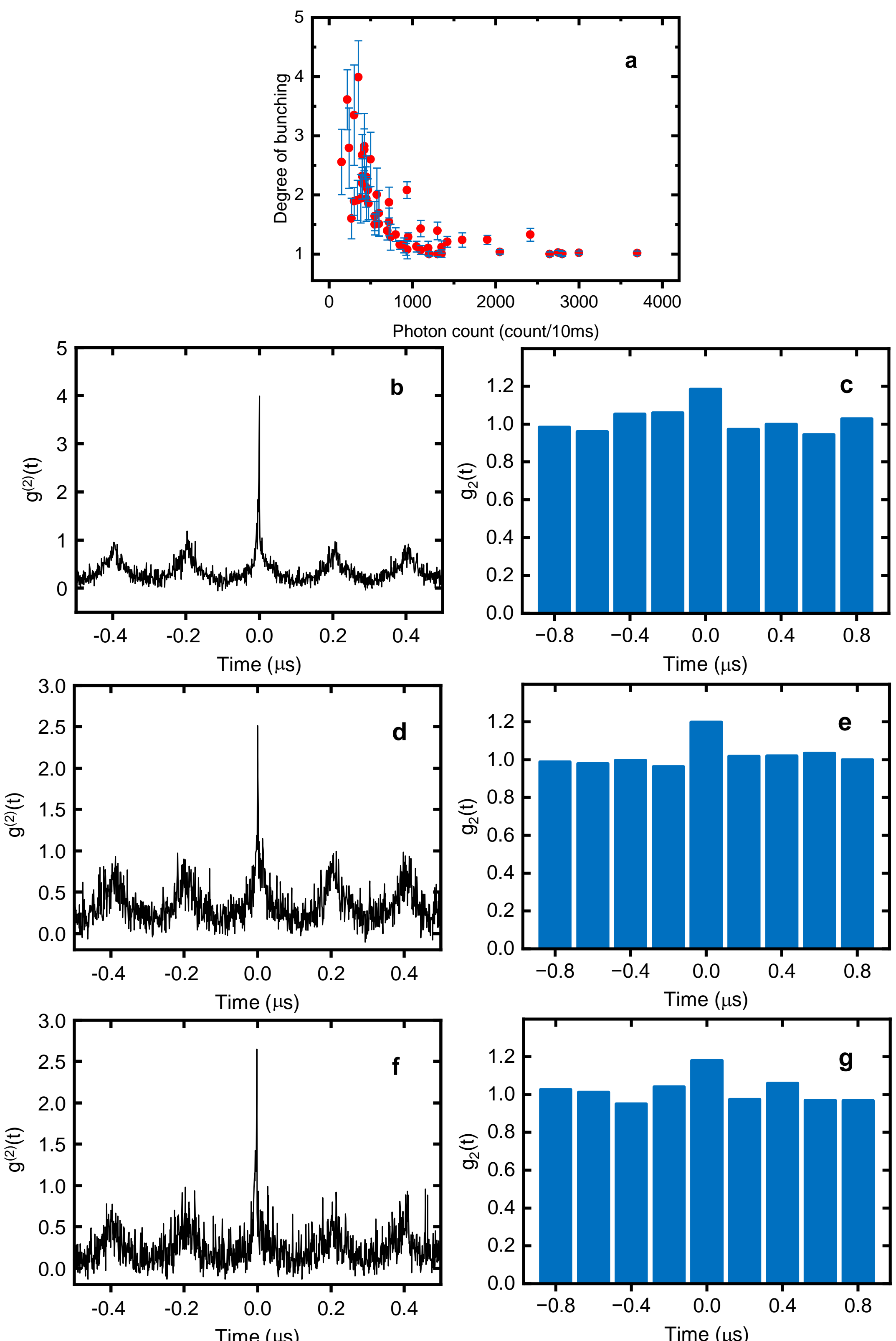

**Supplementary Fig. 8** (a) PL intensity dependent degree of photon bunching obtained with the same excitation power on 49 single superlattices; the vertical bars represent the experimental errors; (b,d,f) Examples of the photon coincidence measurements corresponding the low PL intensity region; (c,e,g) Plots of integrated areas of the $g^{(2)}(t)$ peaks in the respective figures (b,d,f).

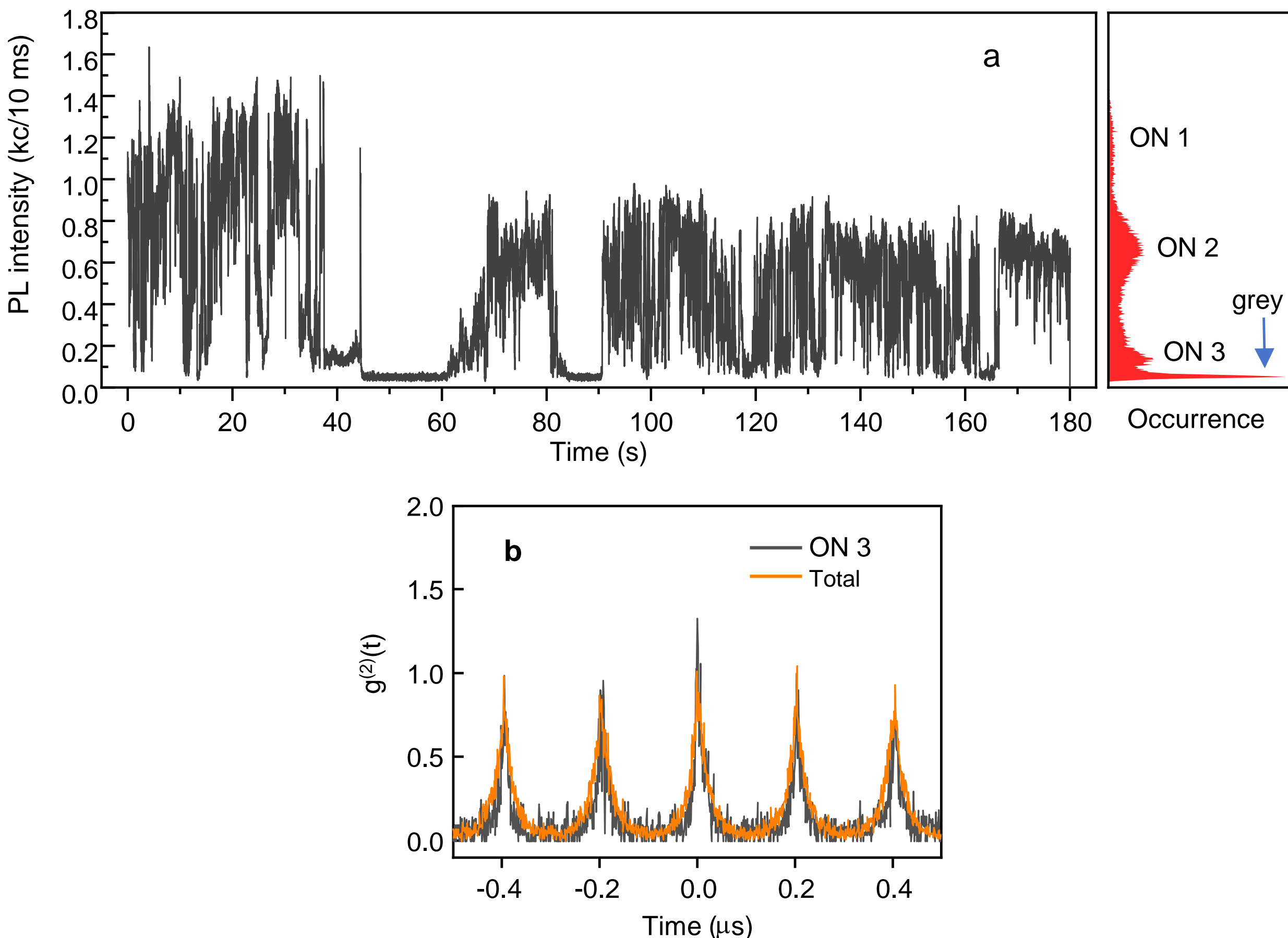

**Supplementary Fig. 9** (a) Left: PL intensity fluctuation of a single superlattice; Right: Histogram of PL intensity levels during the blinking time interval; the three ON-states and a grey-state levels are indicated in the figure; (b) Photon coincidence measurements of emission from the superlattice. Orange: analyzed from the total emission; black: analyzed from the lowest ON 3 state. Only the ON 3 state shows photon bunching. All data were acquired using the APD.

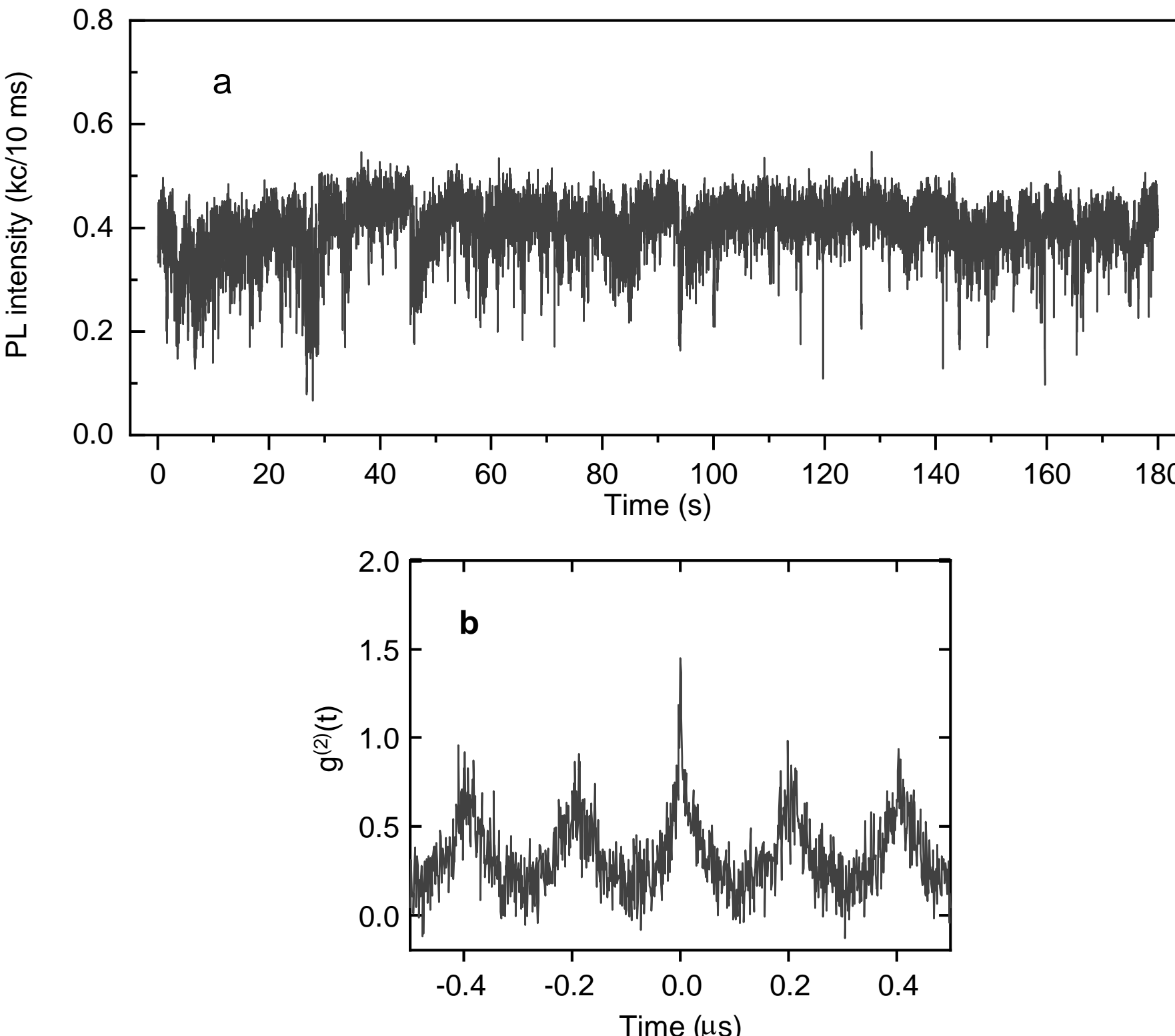

**Supplementary Fig. 10** (a) PL intensity fluctuation of a QD superlattice that did not show clear blinking; (b) Photon coincidence measurement of emission from the non-blinking superlattice showing photon bunching. All data were acquired using the APD.

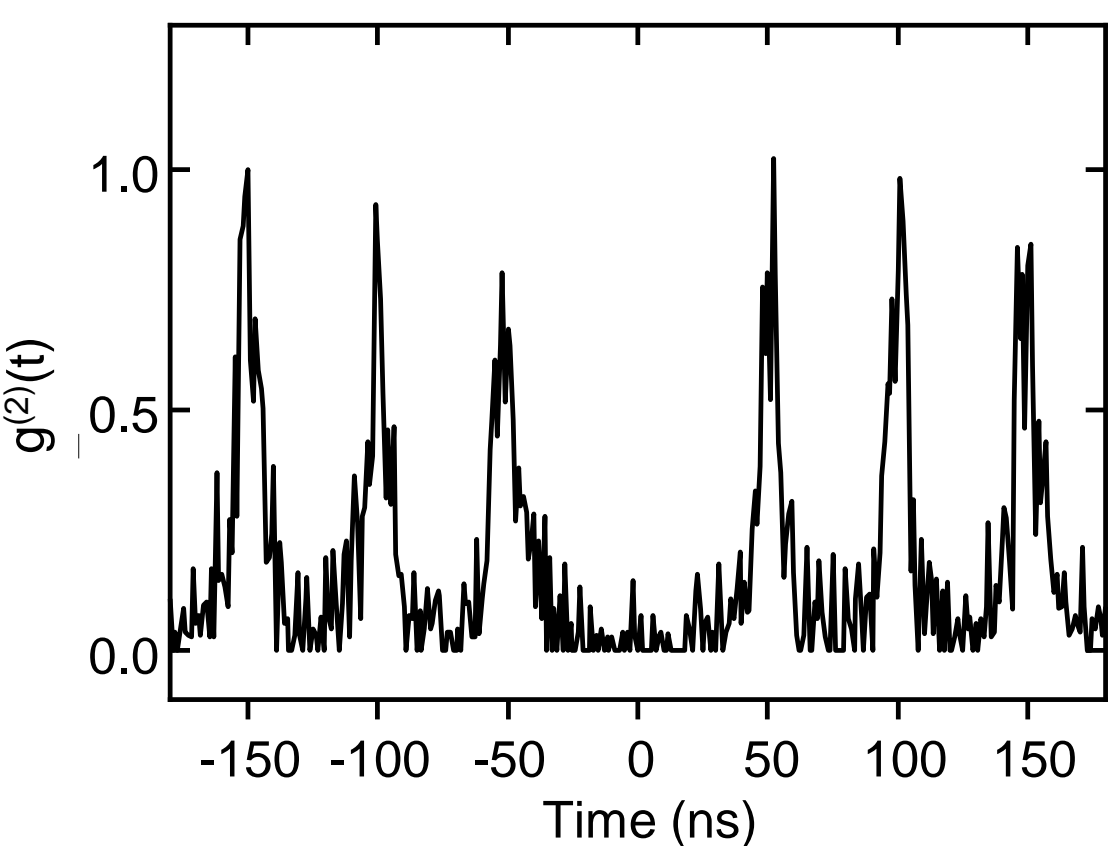

**Supplementary Fig. 11** Photon coincidence measurement of emission from a single $CsPbBr_3$ QD, showing photon anti-bunching as expected for a single quantum emitter.

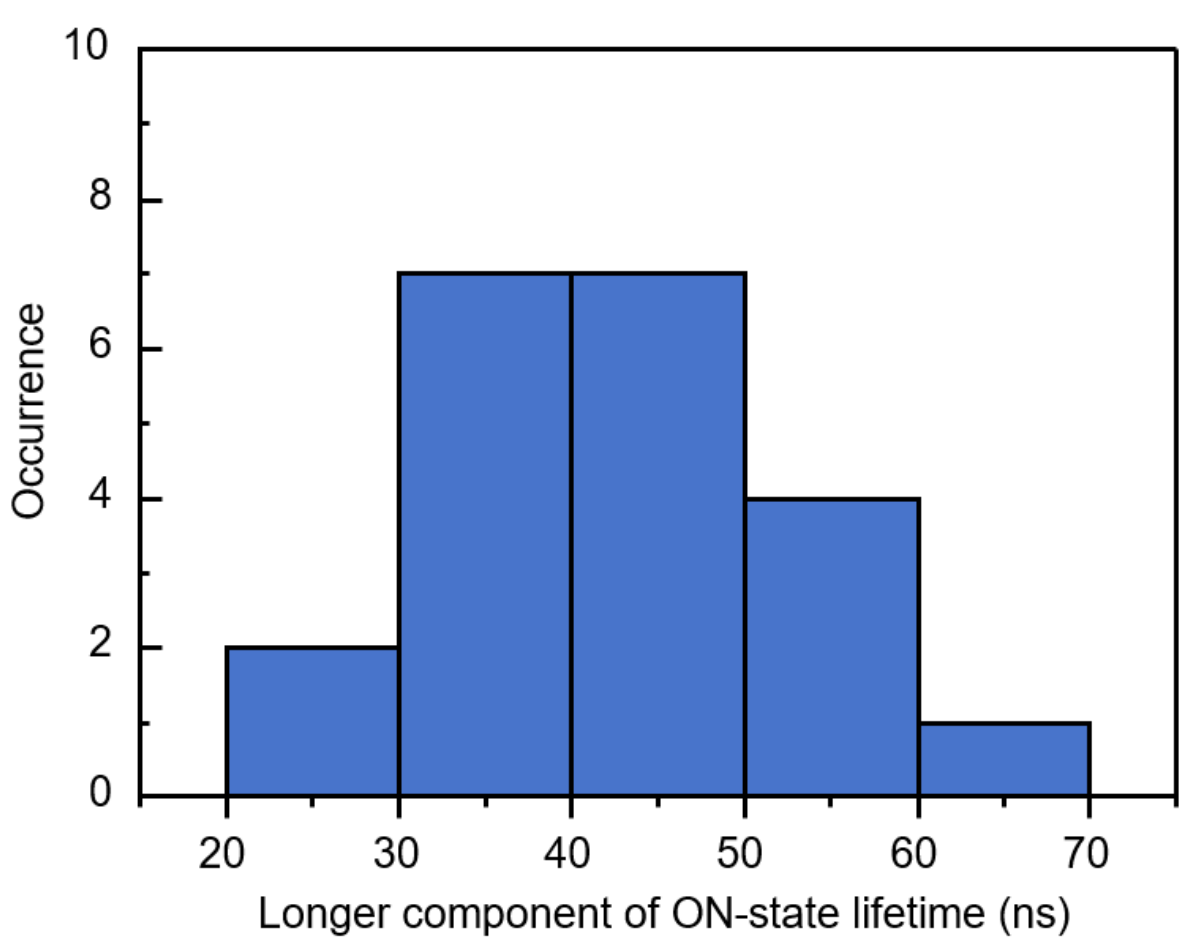

**Supplementary Fig. 12** histogram of the longer component of ON-state lifetimes measured on 21 individual superlattices.

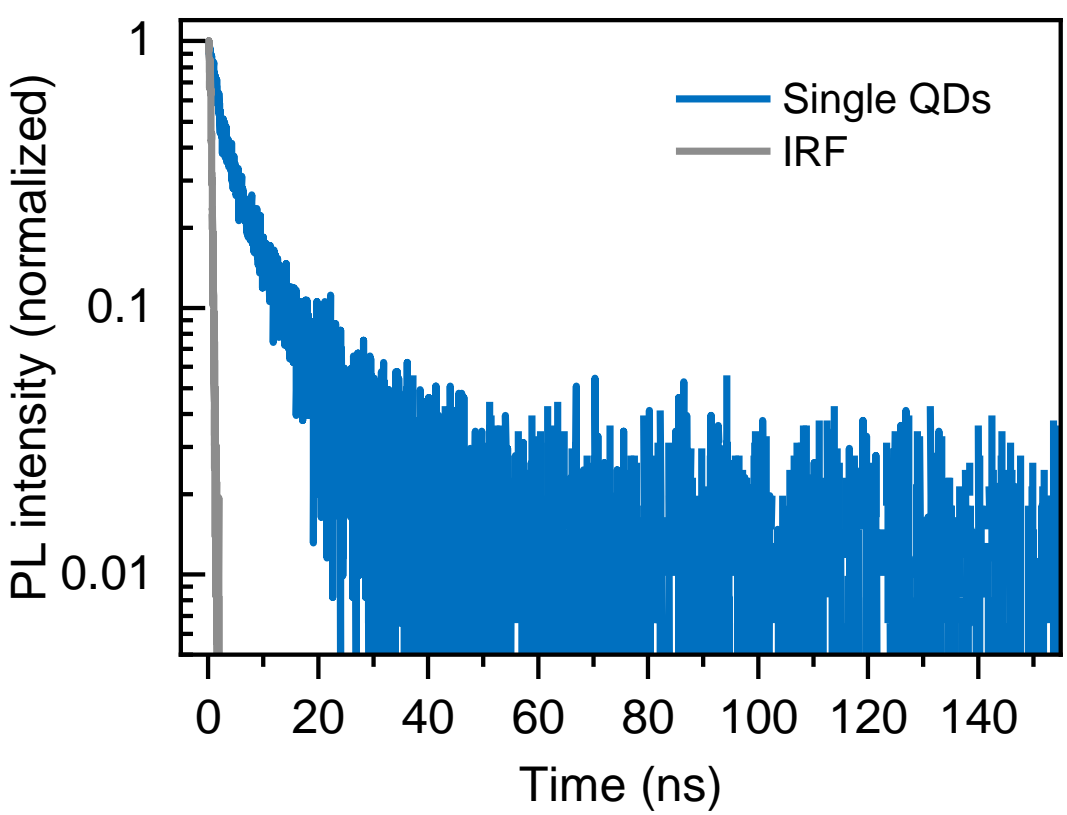

**Supplementary Fig. 13** PL lifetime measured on a random ensemble of isolated QDs. Two-exponential fitting of the data yield lifetime components of 2.4 ns and 12 ns.

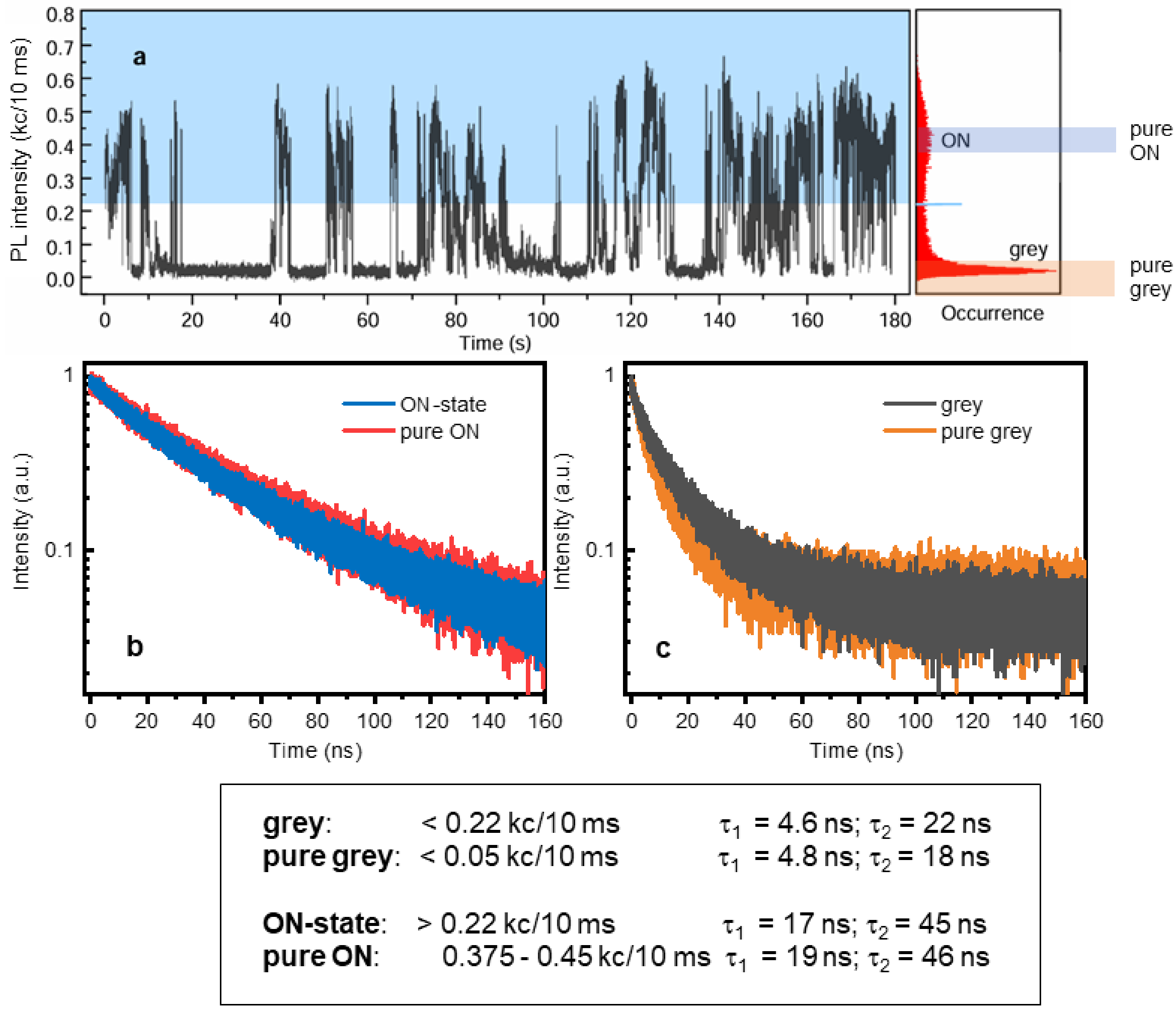

| | | |
|---|---|---|
| **grey**: | < 0.22 kc/10 ms | $\tau_1$ = 4.6 ns; $\tau_2$ = 22 ns |
| **pure grey**: | < 0.05 kc/10 ms | $\tau_1$ = 4.8 ns; $\tau_2$ = 18 ns |
| **ON-state**: | > 0.22 kc/10 ms | $\tau_1$ = 17 ns; $\tau_2$ = 45 ns |
| **pure ON**: | 0.375 - 0.45 kc/10 ms | $\tau_1$ = 19 ns; $\tau_2$ = 46 ns |

**Supplementary Fig. 14** (a) Left: PL intensity fluctuation of a $CsPbBr_3$ superlattice; the blue shaded area indicates the PL intensity range of the ON-state (> 0.22 kc/10 ms); Right: Histogram of PL intensity levels corresponding to the time trace; the rectangles indicate the intensity ranges of the pure ON (0.375 - 0.45 kc/10 ms) and pure grey (< 0.05 kc/10 ms) states. (b) PL lifetime of the ON-state (blue) and the pure ON-state (red) of the single superlattice in (a). (c) PL lifetime of the grey state (grey) and the pure grey state (orange) of the single superlattice in (a). The table below the graphs summarizes lifetime components obtained by two-exponential fitting of the PL decays.

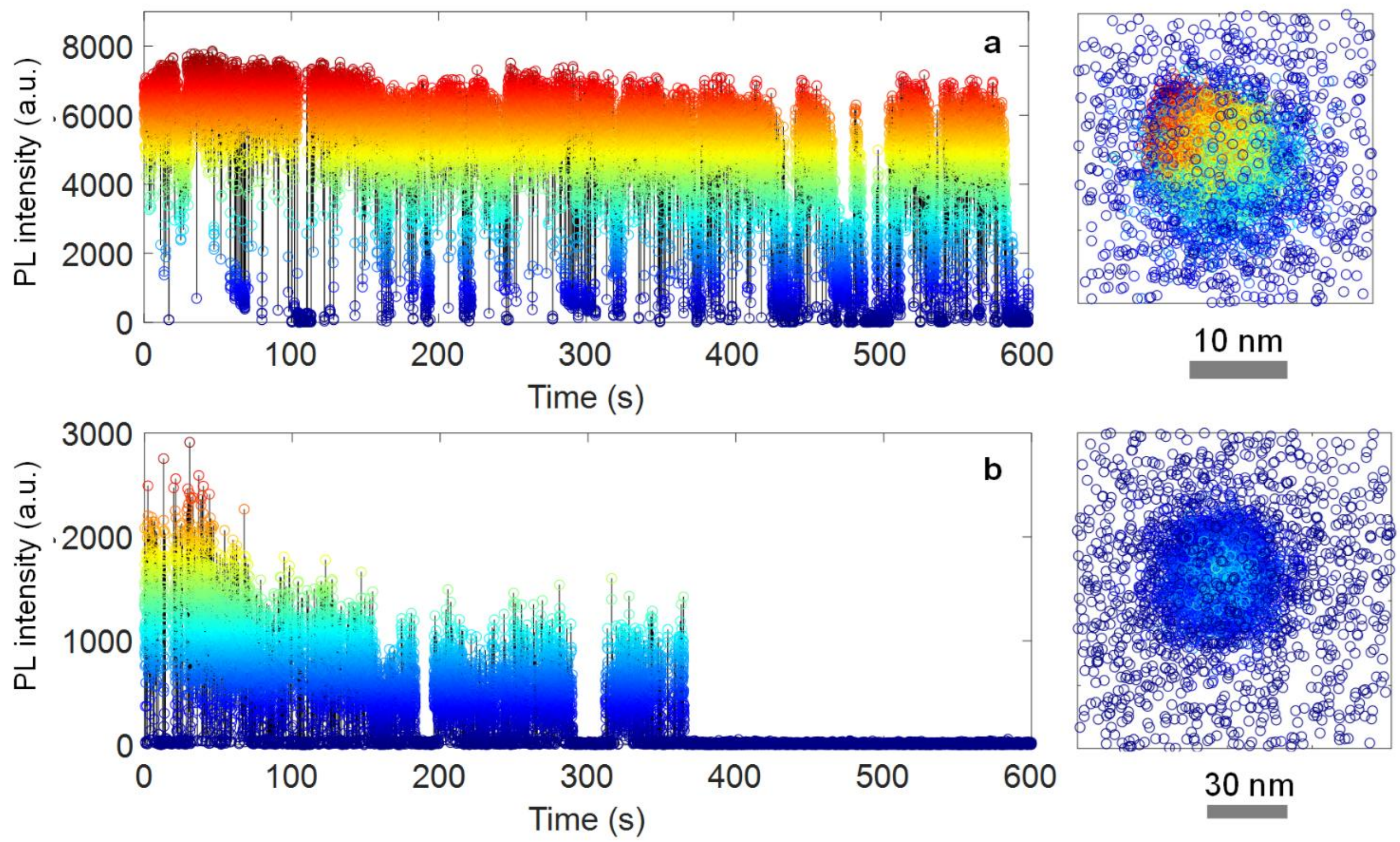

**Supplementary Fig. 15** (left) PL intensity fluctuation of $CsPbBr_3$ superlattice; different PL intensity levels are represented by distinct colors, with the highest intensity marked in red and the lowest in blue; (right) Localized positions of PL corresponding to different intensity levels. (a) Representative blinking trace; (b) Blinking trace that underwent partial photobleaching. The data were acquired using EM-CCD.

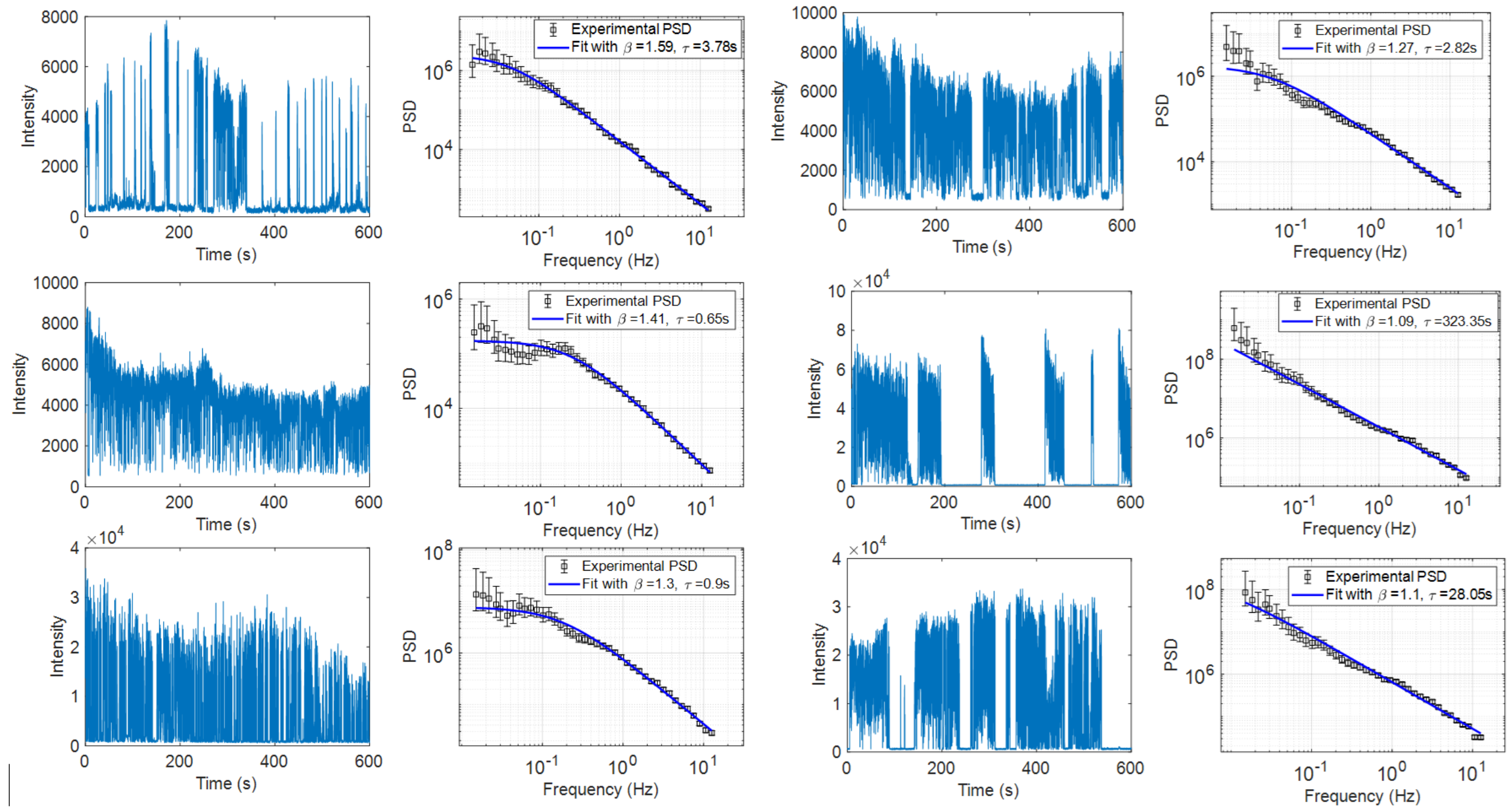

**Supplementary Fig. 16** Examples of blinking traces (left columns, blue lines) and power spectral density (PSD) analyses (right columns) of individual QD superlattices. The insets show parameters obtained by fitting the PSDs using the Eq. (1). The data were acquired using EM-CCD.

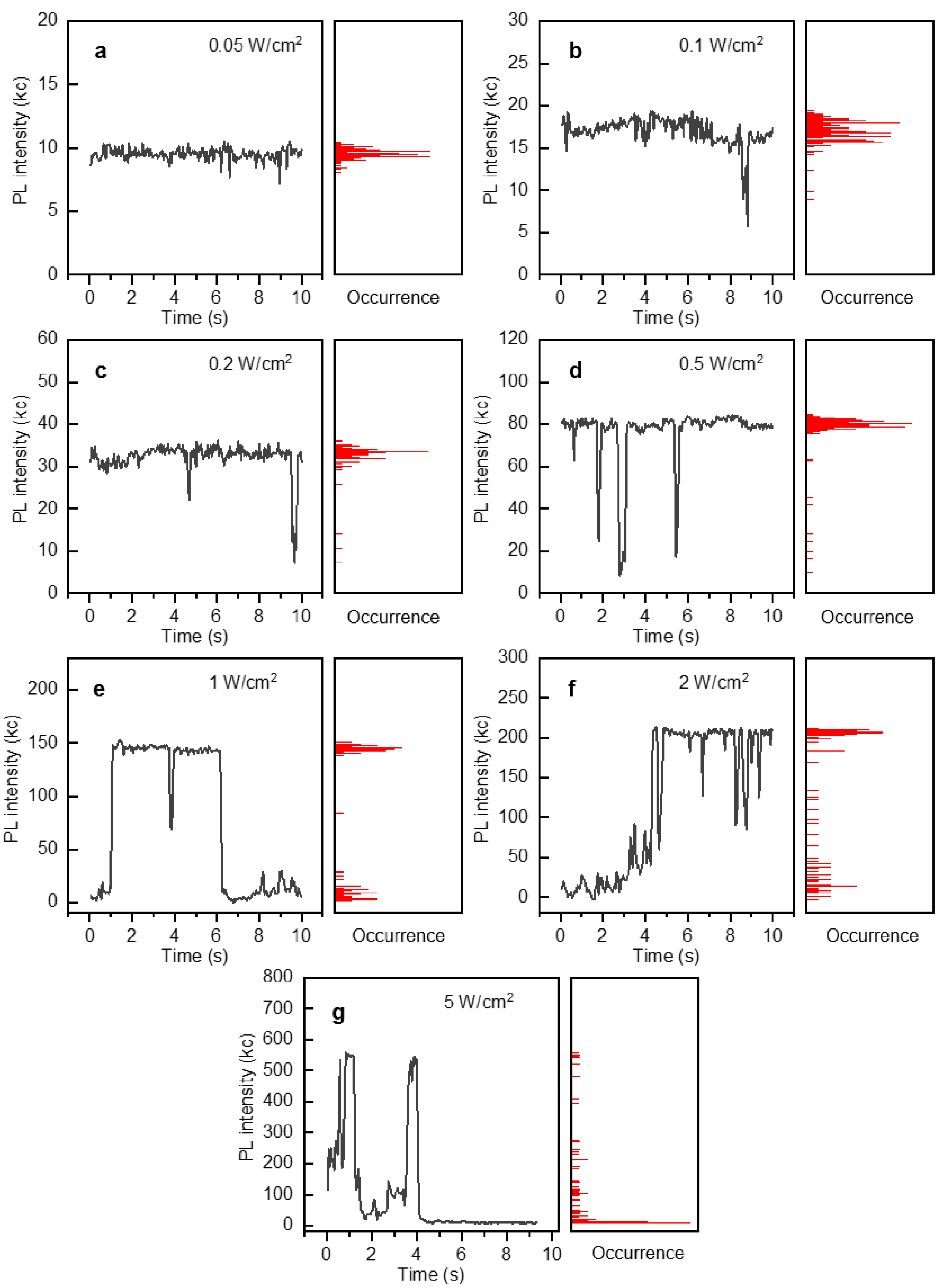

**Supplementary Fig. 17** PL linking traces taken at different excitation powers of (a) 0.05 W/cm$^2$; (b) 0.1 W/cm$^2$; (c) 0.2 W/cm$^2$; (d) 0.5 W/cm$^2$; (e) 1.0 W/cm$^2$; (f) 2.0 W/cm$^2$; (g) 5.0 W/cm$^2$. The data are taken with EM-CCD camera with integration time of 50 ms.

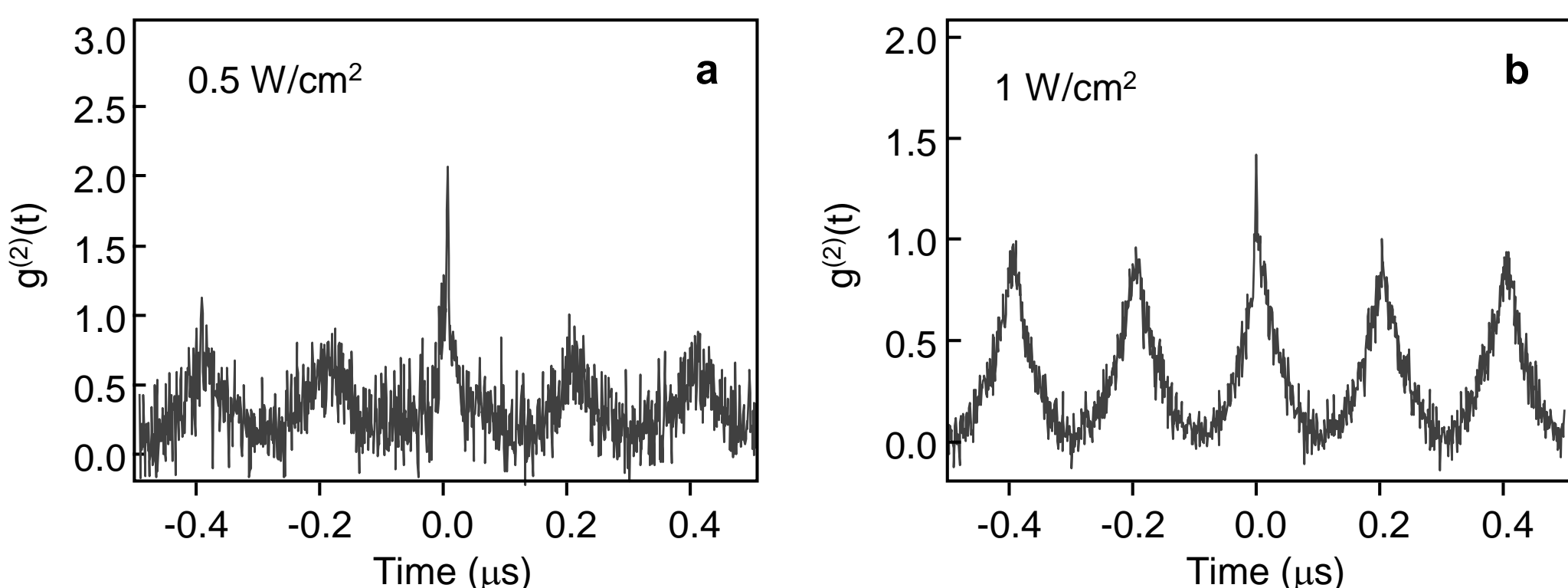

**Supplementary Fig. 18** (b) Photon coincidence measurement of emission from the same superlattice at different excitation powers; (a) 0.5 W/cm$^2$, photon bunching degree of 2.07; (b) 1.0 W/cm$^2$, photon bunching degree of 1.42;

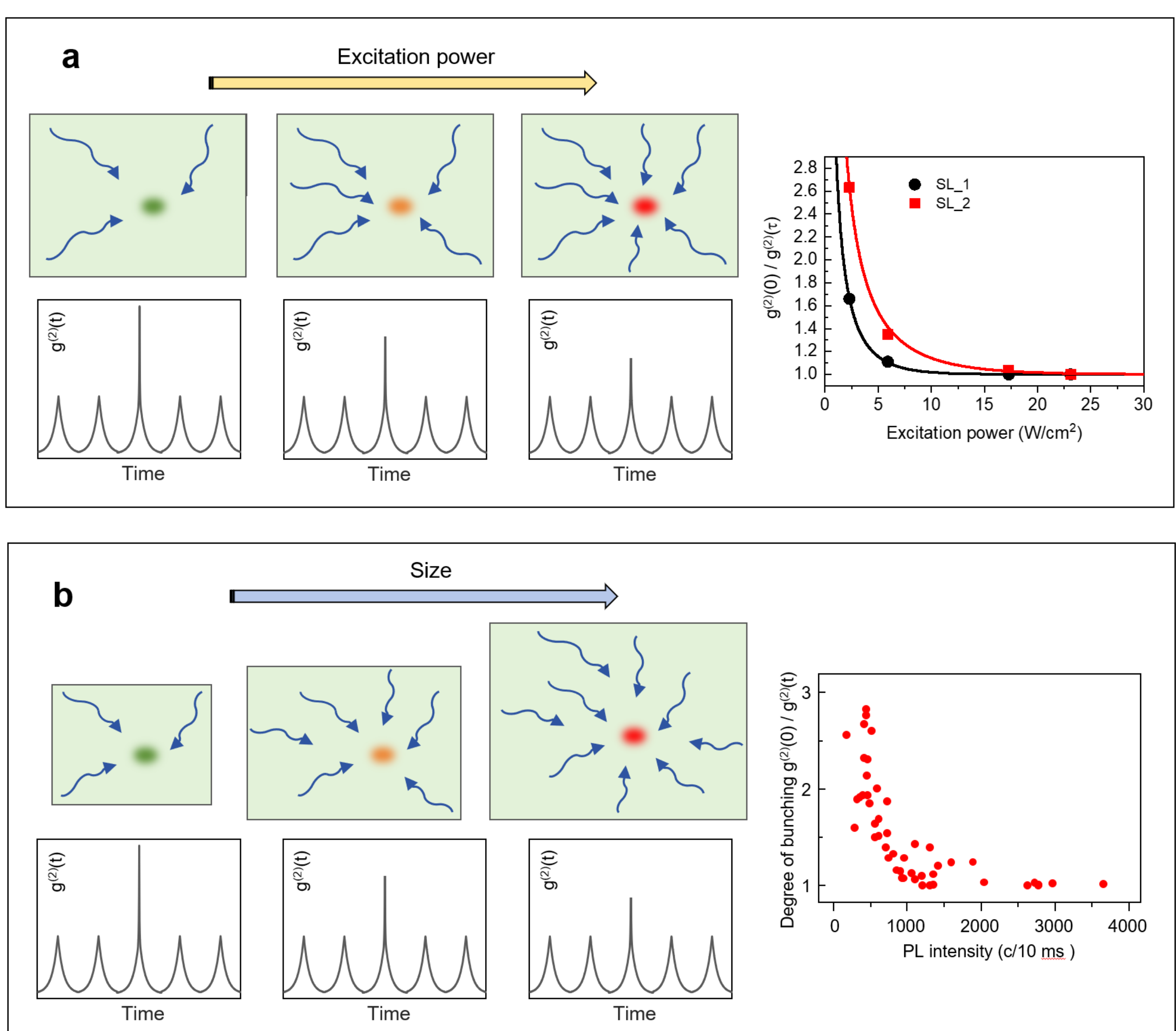

**Supplementary Fig. 19** Schematic explanation of the decrease of the degree of bunching with increasing excitation power (a) and with increasing superlattice size (b). In both cases, the increased exciton density in the emission center leads to lower bunching degree, following the Eq. (3).

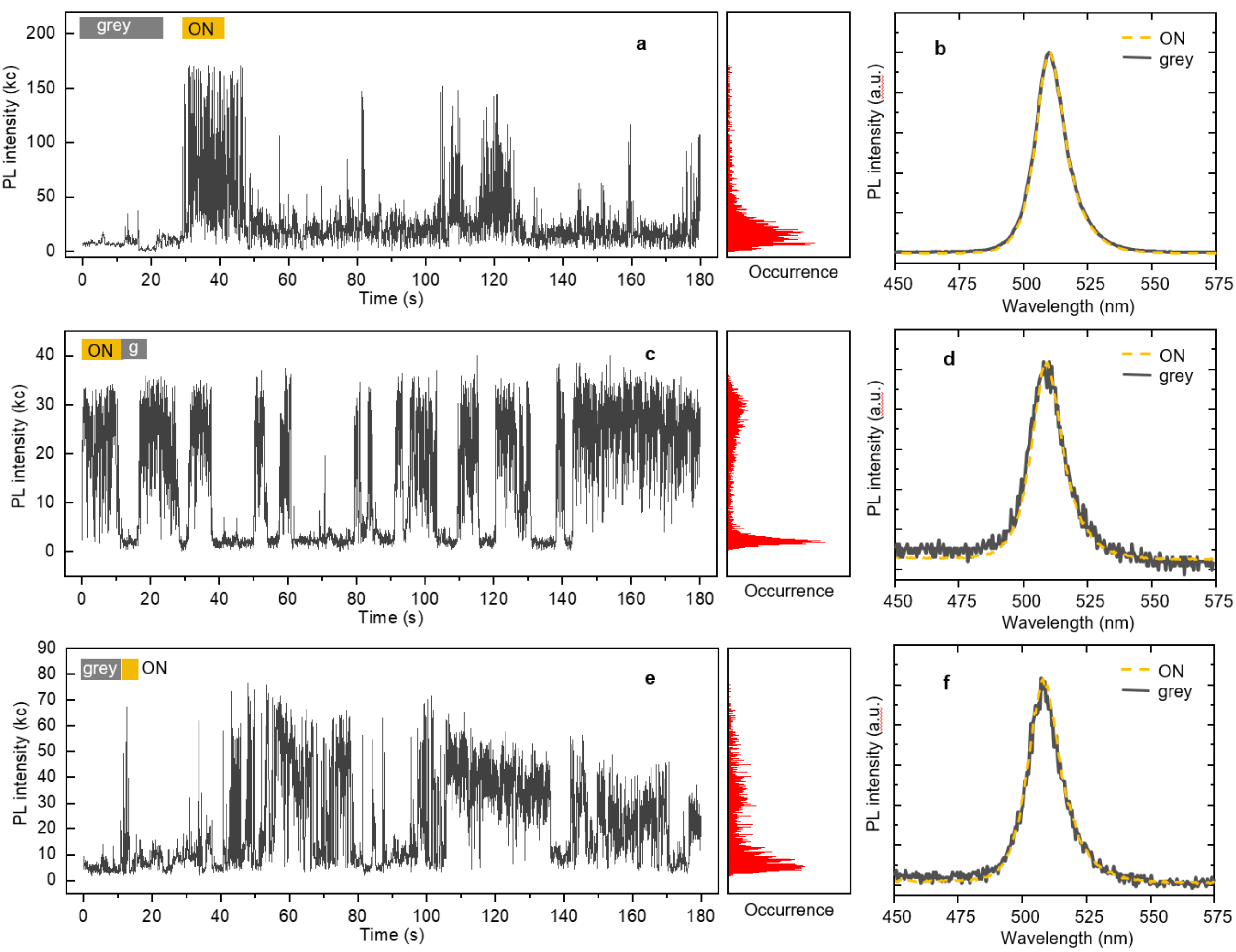

**Supplementary Fig. 20** (a), (c), (e) PL intensity blinking traces of single superlattices; the grey and yellow rectangles above the frame indicate time intervals over which the grey state and ON state PL spectra were integrated. (b), (d), (f) PL spectra integrated from the intervals indicated on the left; dashed yellow - ON state, full grey - grey state. The data were acquired using EM-CCD.

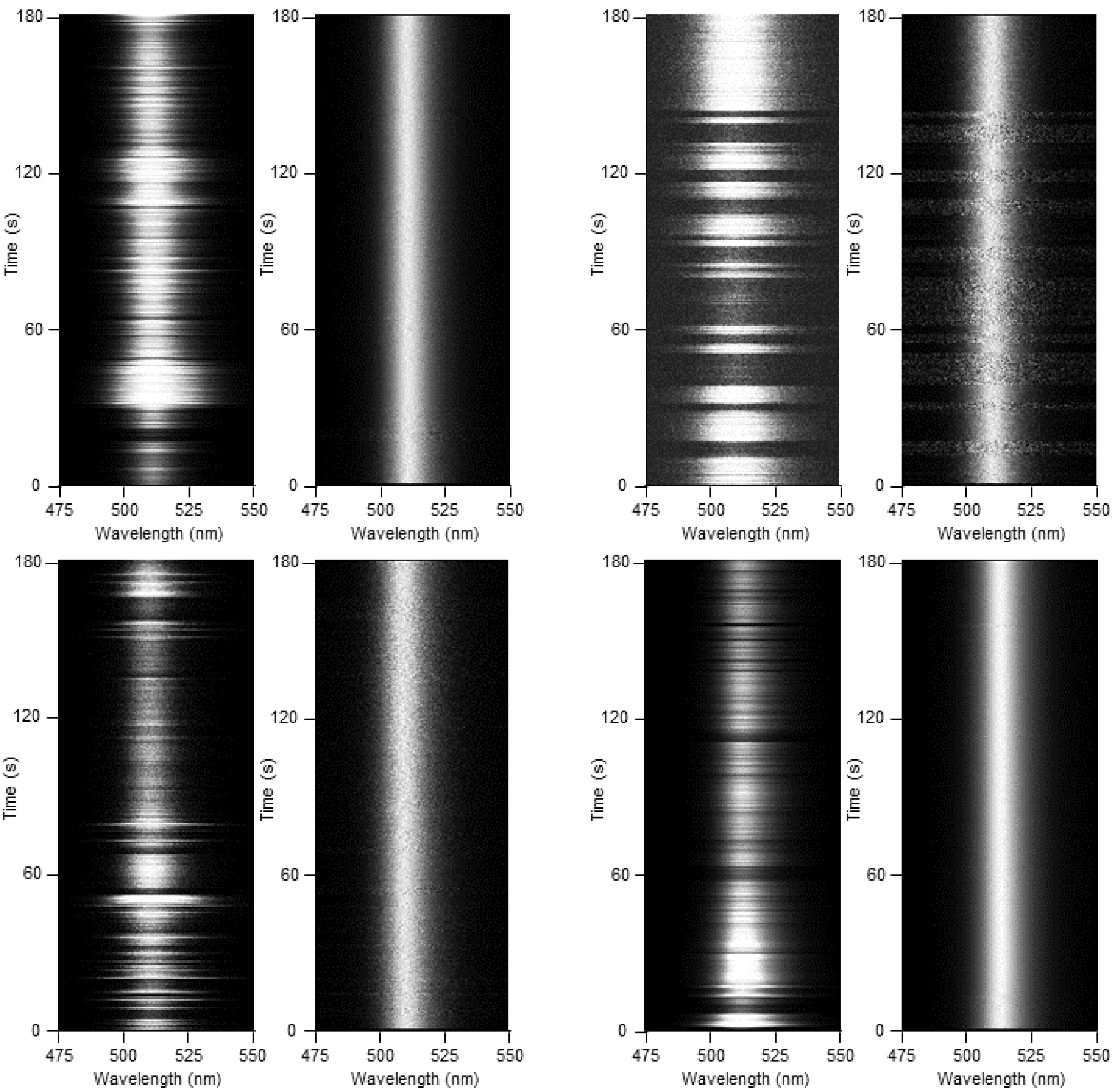

**Supplementary Fig. 21** 2D plots of a time evolution of PL spectra measured from four different single superlattices. The relative PL intensity is represented by the grey scale. The right panels show the spectra in normalized form.

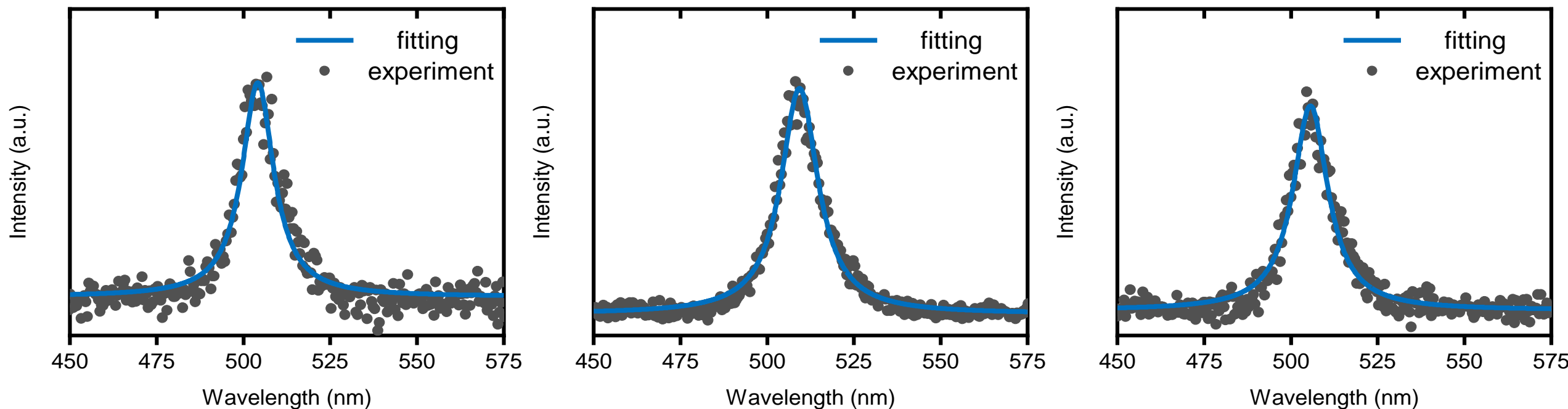

**Supplementary Fig. 22** Three examples of PL spectra of single $CsPbBr_3$ QDs (black symbols) fitted with Lorentzian lineshape (blue lines)

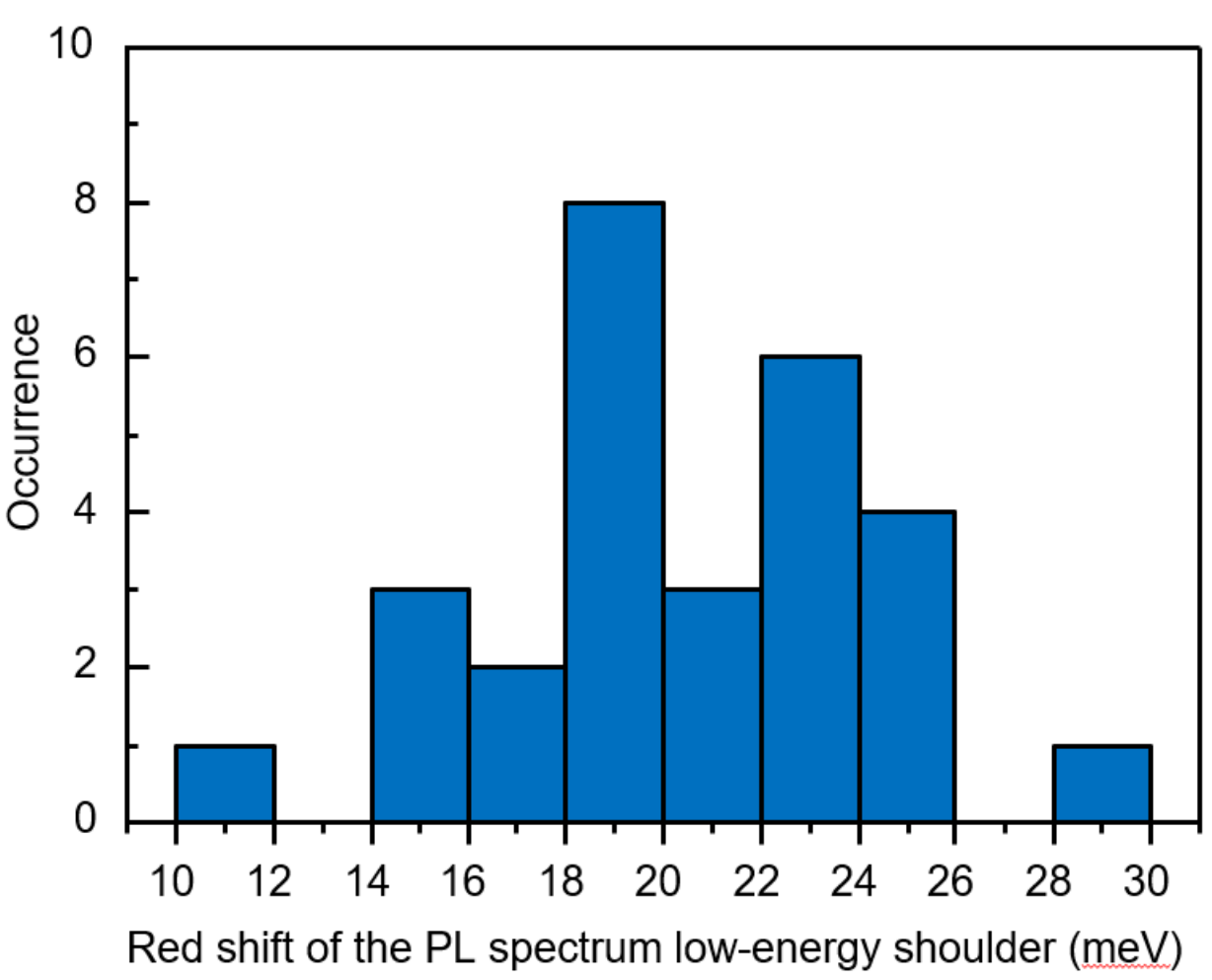

**Supplementary Fig. 23** Histogram of the energy difference between the two Gaussian components of the deconvolved PL spectra measured for 28 individual superlattices.

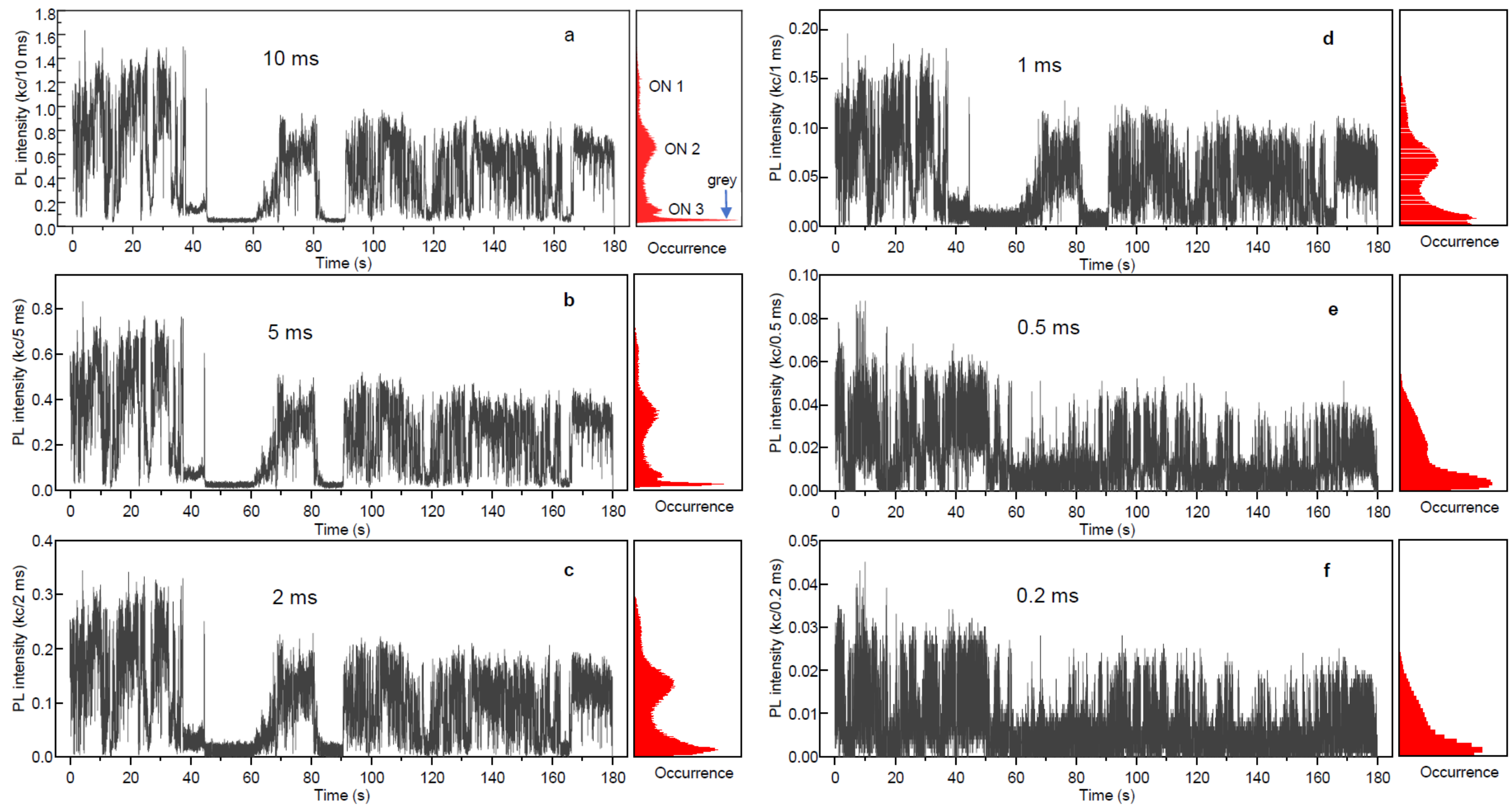

**Supplementary Fig. 24** PL intensity fluctuation of a single blinking superlattice (shown originally in **Supplementary Fig. 9**) measured with APD using the TTTR mode and analyzed with different binning times: (a) 10 ms (same as in **Supplementary Fig. 9**); (b) 5 ms; (c) 2 ms; (d) 1 ms; (e) 0.5 ms; (f) 0.2 ms. The right parts of the graphs are the corresponding intensity histograms. With the shorter times (< 2 ms), the individual intensity levels in the histogram get smeared out due to the increased noise. At the same time shortening of the integration time did not reveal any additional features such as unresolved OFF or grey states.

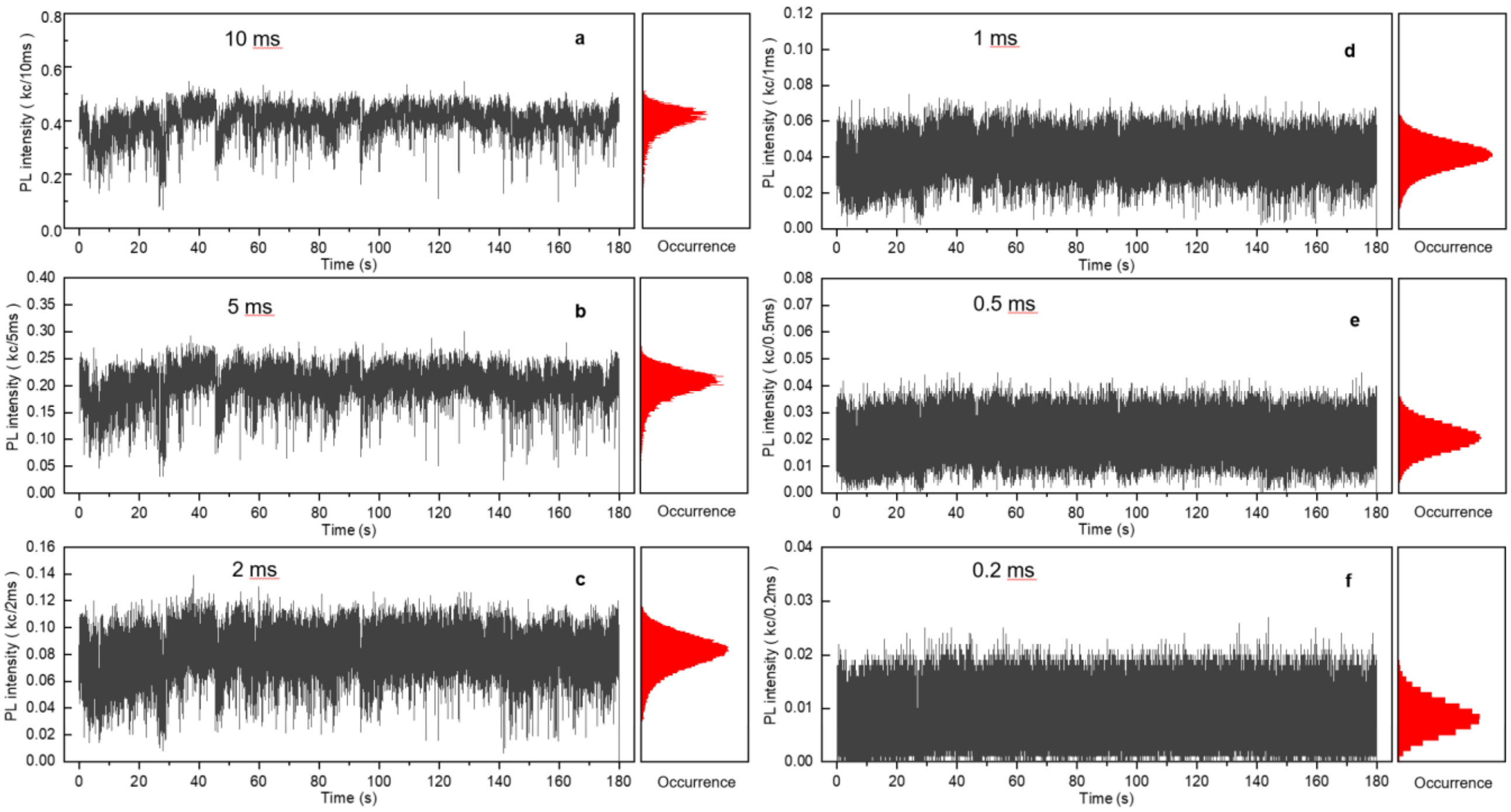

**Supplementary Fig. 25** PL intensity fluctuation of a single non-blinking superlattice (shown originally in **Supplementary Fig. 10**) measured with APD using the TTTR mode and analyzed with different binning times: (a) 10 ms (same as in **Supplementary Fig. 10**); (b) 5 ms; (c) 2 ms; (d) 1 ms; (e) 0.5 ms; (f) 0.2 ms. The right parts of the graphs are the corresponding intensity histograms. With the shorter times (< 2 ms), the individual intensity levels in the histogram get smeared out due to the increased noise. At the same time shortening of the integration time did not reveal any additional features such as unresolved OFF or grey states.